\documentclass[a4paper,11pt]{article}

\usepackage[left=2.5cm,right=2.5cm,top=2.5cm,bottom=2.5cm]{geometry}
\usepackage{graphicx,amssymb,amsmath,amsthm,mathrsfs,setspace,subcaption,cite,authblk,float,upgreek}
\usepackage[all]{xy}

\usepackage[colorlinks=true,bookmarks=true]{hyperref}
\usepackage{breakurl} 

\usepackage{mathtools}

\DeclarePairedDelimiterX\braket[2]{\langle}{\rangle}{#1 \delimsize\vert #2}

\theoremstyle{definition}

\hypersetup{citecolor=blue}
\newcommand{\dif}{\mathrm{d}}
\newcommand{\Eqref}[1]{(\ref{#1})}
\newcommand{\half}{\frac{1}{2}}

\newcommand{\brac}[1]{\left(#1 \right)}
\newcommand{\sbrac}[1]{\left[#1\right]}

\numberwithin{equation}{section}
 
\begin{document}

\title{Structure of test magnetic fields and charged particle motion around the Hayward spacetime}

\author{Ziou Yang\footnote{Email: phy2009487@xmu.edu.my}\;}
\author{Yen-Kheng Lim\footnote{Email: yenkheng.lim@gmail.com, yenkheng.lim@xmu.edu.my}}

\affil{\normalsize{\textit{Department of Physics, Xiamen University Malaysia, 43900 Sepang, Malaysia}}}

\date{\normalsize{\today}}
\maketitle

\renewcommand\Authands{ and }
\begin{abstract}
 A configuration of a test magnetic field in Hayward spacetime is obtained by solving Maxwell's equation with the Hayward metric as the background. The magnetic field lines show a dipole loop-like configuration in the regular Hayward interior, and tends to an asymptotically uniform structure away from the cylindrical axis. The motion of charged particles is then studied in this spacetime. The parameters and stability of circular orbits on the equatorial plane are studied. Aspects of non-equatorial motion are also studied.
\end{abstract}

\section{Introduction} \label{intro}

The presence of curvature singularities is a typically undesirable properties of a particular spacetime, though they seem to be an unavoidable feature of black holes. However, there are proposed solutions \cite{Bardeen,Dymnikova:1992ux,Ayon-Beato:1999qin} which describe black holes or gravitating bodies which are regular. That is, they do not carry curvature singularities.  (See Ref.~\cite{Ansoldi:2008jw} for a review about regular black holes.)

In this paper, we are specifically interested in the one derived by Hayward \cite{Hayward:2005gi}. Hayward's solution is a static and spherically symmetric spacetime which describes either a black hole or a horizonless gravitating object, depending on the parameters of the solution. The solution is non-vacuum and requires a particular form for the stress tensor, though an explicit model can be provided by a non-linear electrodynamics source \cite{Kruglov2015,Ali:2019bcn,Mazharimousavi:2019jja,Kruglov2021,IlichKruglov:2021pdw}. The vicinity around the origin of this spacetime is well-approximated by the de Sitter metric, and is completely regular at $r=0$. Far away from the origin, the geometry tends to the asymptotically-flat, Schwarzschild-like geometry.

Various aspects of the Hayward solution has been studied and extended over the years \cite{Flachi:2012nv,Halilsoy:2013iza,Pourhassan:2016qoz,Um:2019qgc,NaveenaKumara:2020lgq}. Of particular relevance to this paper is the study of particle motion in this spacetime. Null and time-like geodesic motion in the Hayward spacetime has been worked out in Refs.~\cite{Abbas:2014oua,Schee:2015nua,Chiba:2017nml,Hu:2018old}. The Hayward spacetime has been generalised to the rotating case in \cite{Bambi:2013ufa,Torres:2016pgk}. The motion of particles and light in these rotating counterparts have also been studied in Refs.~\cite{Bautista-Olvera:2019blb,Amir:2015pja,Kumar:2019pjp}. 

If we were to consider the Hayward spacetime as a possible astrophysical object, it would be reasonable to include the presence of magnetic fields to the solution, since magnetic fields are expected to influence the physics around black holes \cite{Aliev:1986wu,Aliev:1989wx,Aliev:2002nw,Frolov:2010mi,Frolov:2011ea,Zahrani:2013up,Shiose:2014bqa,Kolos:2015iva,Shoom:2015uba}. Beyond astrophysical phenomena, magnetic fields also play important roles in other theoretical contexts. As such the inclusion of (electro-)magnetic fields in various spacetimes has been studied \cite{Herdeiro:2015vaa,Costa:2015gol,Vesely:2019ajp,Vesely:2021jlc}.

It was argued in Refs.~\cite{Aliev:1989wx,Frolov:2010mi} that magnetic fields generated in astrophysical phenomena do not appreciably deform the spacetime curvature, and that it suffices to consider test magnetic fields. As such, one of the aims of this paper is to immerse the Hayward spacetime in a test magnetic field. Interestingly, we found that the magnetic field lines near the de Sitter-like core has the loop structure characteristic of an isolated magnetic dipole. In the Schwarzschild-like region the magnetic field tends to the uniform configuration which agrees with Wald's solution for test magnetic field around the pure Schwarzschild black hole.

The second main aim of this paper is to study the motion of charged particles in this spacetime. In the Schwarzschild-like region where the magnetic field is approximately uniform, the behaviour is largely similar to the pure Schwarzschild case studied in \cite{Frolov:2010mi,Kolos:2015iva}. In particular, there are also `curly' orbits caused by the combination of the magnetic Lorentz forces and the radial gravitational forces which causes the azimuthal angular velocity to change signs at various points. More interesting types of trajectories can be found for the horizonless Hayward spacetime, where the de Sitter-like core is accessible. For instance, there are two potential wells on the $z$-axis where particles can be trapped. Because of the loop structure of the magnetic field lines in this region, the particle trajectory exhibits `bounce' and `drift' motion similar to particles around planetary magnetic fields. 

The rest of this paper is organised as follows. In Sec.~\ref{sec_Maxwell}, we solve Maxwell's equation under the Hayward metric to obtain the test magnetic field. The equations of motion for a charged particle in this spacetime is derived in Sec.~\ref{sec_EOM}. In Sec.~\ref{sec_Equatorial}, we consider trajectories confined to the equator, and non-equatorial motion will be considered in Sec.~\ref{sec_NonEquatorial}. Conclusions and closing remarks will be given in Sec.~\ref{sec_conclusion}. We will be working in geometric units where $c=G=1$, where $c$ is the speed of light and $G$ is Newton's constant. Our convention for Lorentzian signature is $(-,+,+,+)$.

\section{Test magnetic field in the Hayward spacetime} \label{sec_Maxwell}

We begin with a brief review of the Hayward solution \cite{Hayward:2005gi}. Its spacetime metric is given by
\begin{subequations}\label{Hayward_metric}
\begin{align}
 \dif s^2&=-f(r)\dif t^2+\frac{\dif r^2}{f(r)}+r^2\dif\theta^2+r^2\sin^2\theta\,\dif\phi^2,\\
 f(r)&=1-\frac{2mr^2}{r^3+2m\ell^2},
\end{align}
\end{subequations}
where $m$ is the \emph{mass parameter} and $\ell$ the \emph{Hayward parameter} of the spacetime. The notable feature of this solution is the absence of a curvature singularity at the origin $r\rightarrow0$ for $\ell\neq0$. This spacetime is a solution to the Einstien equation $R_{\mu\nu}-\half R g_{\mu\nu}=8\pi T_{\mu\nu}$, where the energy density $\rho=-{T^t}_t$, radial pressure $p={T^r}_r$, and tangential pressures $p_\perp={T^\theta}_\theta={T^\phi}_\phi$ are given by 
\begin{align}
 \rho=-p=\frac{3m^2\ell^2}{\pi \brac{r^3+2m\ell^2}^2},\quad p_\perp=\frac{3m^2\ell^2\brac{r^3-m\ell^2}}{\pi \brac{r^3+2m\ell^2}^3}. \label{stress}
\end{align}
Horizons are characterised by the roots of $f$. For values of $\ell$ satisfying $\frac{\ell}{m}<\frac{4}{3\sqrt{3}}\simeq0.7698$, $f$ has two distinct positive roots $r_\pm$ with the notation assigned to the order $r_-<r_+$. We refer to this case as the \emph{Hayward black hole} with $r_+$ being the outer horizon and $r_-$ the inner horizon. The asymptotically flat exterior static domain is $r>r_+$. There is another static domain interior to the inner horizon with $0<r<r_-$. If $\frac{\ell}{m}>\frac{4}{3\sqrt{3}}$, $f$ has no real positive roots and the spacetime is static all throughout $0<r<\infty$, and we refer to this case as the \emph{horizonless Hayward spacetime}. The critical case $\frac{\ell}{m}=\frac{4}{3\sqrt{3}}$ is the where $f$ has a degenerate horizon.

In the study of test magnetic fields in the Hayward spacetime, it will be convenient to define a characteristic radius 
\begin{align}
 r_c=\brac{4m\ell^2}^{1/3}. \label{rc_def}
\end{align}
The significance of $r_c$ is that $f'(r)<0$ for $r<r_c$ and $f'(r)>0$ for $r>r_c$. Roughly speaking, it can be seen that the metric is approximately described by de Sitter space for $r\ll r_c$ (and is regular at $r=0$), while the metric tends to the asymptotically flat Schwarzschild-like geometry for $r\gg r_c$.

We now turn to the task of immersing the spacetime in a test magnetic field. To do this we solve the Maxwell equation $\nabla_\lambda F^{\lambda\nu}=4\pi J^\nu$, where $F=\dif A$ is the exterior derivative of the gauge potential $A$ and $J^\nu$ is the electromagnetic current four-vector. The strength of $F$ is assumed to be sufficiently small such that it does not backreact into the spactime curvature.

If we had a \emph{vacuum} solution of Einstein's equation, a quick way to include a test magnetic field is to observe that Maxwell's equation expressed in terms of $A$ is equivalent to Killing's equation under a certain gauge \cite{Wald:1974np} so that any Killing vector of the spacetime can be the gauge potential for the test magnetic field. Unfortunately, this trick is not available for non-vacuum spacetimes, as in our present situation. Therefore we simply solve Maxwell's equation directly. Test magnetic fields for spacetimes with cosmological constant  \cite{Herdeiro:2015vaa,Costa:2015gol} and in non-vacuum black holes in modified gravity \cite{Azreg-Ainou:2016tkt} have previously been obtained this way.

To this end, we consider the following ansatz for the gauge potential:
\begin{align}
 A=\Psi(r,\theta)\;\dif\phi,
\end{align}
where the function $\Psi(r,\theta)$ does not depend on $\phi$ and $t$. Under this ansatz, along with $J^\mu=(0,0,0,J^\phi)$, the Maxwell equation becomes 
\begin{align}
 r^2\partial_r\brac{f\partial_r\Psi}+\sin\theta\partial_\theta\brac{\frac{1}{\sin\theta}\partial_\theta\Psi}=4\pi r^4\sin^2\theta J^\phi. \label{PDE}
\end{align}
In the homogeneous case ($J^\phi=0$), we solve \Eqref{PDE} by applying a separation of variables:
\begin{align*}
 \Psi(r,\theta)=-r^2\chi(r)\sin\theta\frac{\dif\Theta(\theta)}{\dif\theta},
\end{align*}
where $\chi(r)$ and $\Theta(\theta)$ are functions of $r$ and $\theta$, respectively. Then we find that there exists a separation constant $n(n+1)$ which splits Eq.~\Eqref{PDE} into two ordinary differential equations:
\begin{subequations}\label{ODE}
\begin{align}
 \frac{\dif}{\dif r}\brac{f\frac{\dif}{\dif r}\brac{r^2\chi}}&=n(n+1)\chi,\label{chi_eqn}\\
 \frac{1}{\sin\theta}\frac{\dif}{\dif\theta}\brac{\sin\theta\frac{\dif\Theta}{\dif\theta}}&=-n(n+1)\Theta. \label{Theta_eqn}
\end{align}
\end{subequations}
Note that Eq.~\Eqref{Theta_eqn} is simply Legendre's equation. Therefore the solution is simply 
\begin{align}
 \Theta(\theta)=P_n(\cos\theta),
\end{align}
where $P_n(x)$ is the $n$-th Legendre polynomial. 

In the present paper, we are mainly interested in generalising the weakly magnetised Schwarzschild solution to the Hayward case. Therefore we consider specifically $n=1$, which is the dipole case. Then Eq.~\Eqref{chi_eqn} is solved by
\begin{align*}
 \chi(r)&=C_1\brac{1-\frac{4m\ell^2}{r^3}}+C_2\brac{1-\frac{4m\ell^2}{r^3}}\int\frac{\dif r}{r^4f(r)\brac{1-\frac{4m\ell^2}{r^3}}^2}, 
\end{align*}
where $C_1$ and $C_2$ are the integration constants. For convenience in our subsequent calculations, we reparametrise our integration constants as $C_1=\frac{B}{2}+CK$ and $C_2=C$, and the homogeneous solution above is written as
\begin{align}
 \chi(r)=\frac{B}{2}U(r)+CU(r)\brac{K+V(r)},\label{chi_gen}
\end{align}
where
\begin{align}
 U(r)=1-\frac{4m\ell^2}{r^3},\quad V(r)=\int\frac{\dif r}{r^4f(r)U(r)^2}.
\end{align}
As the aim of this paper to extend the magnetised Schwarzschild solution to the case of Hayward spacetimes, we are seeking a magnetic field that is asymptotically uniform as $r\rightarrow\infty$. As such one would set $C=0$. However, for the Hayward spacetime where the origin $r=0$ is accessible, the invariant $F^2=F_{\mu\nu}F^{\mu\nu}$ has the behaviour
\begin{align*}
 F^2\simeq\frac{8B^2m^2\ell^4\brac{1+4\cos^2\theta}}{r^6}+\ldots,
\end{align*}
so for this solution, the electromagnetic field strength diverges at the origin. Furthermore, if $F^2$ becomes unboundedly large, then the magnetic field will be strong enough to curve the spacetime and can no longer be considered a test field.

To resolve this singularity, we consider the presence of a current loop lying on the equatorial plane $\theta=\frac{\pi}{2}$ with coordinate $r=a$. The following is an application of the methods of Ref.~\cite{Petterson:1974bt}, adapted to the Hayward spacetime. We take the current four-vector to be $J^\mu=(0,0,0,J^\phi)$, where
\begin{align}
 J^\phi=\frac{If(r)^{1/2}}{r^2}\delta(r-a)\delta(\cos\theta),\label{J_def}
\end{align}
and the delta functions are normalised such that $\int\delta(r-a)\dif r=1$ and $\int\sin\theta\delta(\cos\theta)\dif\theta=1$. The constant $I$ is the current passing through the $r$-$\theta$ plane. To see this, we take $\xi=\brac{r\sin\theta}^{-1}\partial_\phi$ to be the unit vector along $\phi$. The spatial current passing through the plane orthogonal to $\xi$ is 
\begin{align}
 \int J^\phi \xi_\phi\sqrt{g_{\theta\theta}g_{rr}}\;\dif r\dif\theta=\int J^\phi r^2\sin\theta f^{-1/2}\;\dif r\dif\theta=I. 
\end{align}

For $r\neq a$ and $\theta\neq\frac{\pi}{2}$, the current is not present and the homogeneous solution \Eqref{chi_gen} remains valid. With this current loop, we take our solution to be $\Psi=r^2\chi(r)\sin^2\theta$, but with $\chi$ now
\begin{align}
 \chi(r)=\left\{\begin{array}{cc}
                 \chi_1(r), & r>a,\\
                 \chi_2(r), & 0<r<a,
                \end{array}\right.
\end{align}
where $\chi_1(r)=\frac{B}{2}U(r)$ and $\chi_2(r)=CU(r)\brac{K+V(r)}$. The two pieces of the function are to be matched appropriately for the current source $J^\phi$. To achieve this, the solution is required to be continuous at $r=a$, and that Maxwell's equation Eq.~\Eqref{PDE} is to be satisfied at all regions. 

Continuity at $r=a$ requires
\begin{align}
 \lim_{r\rightarrow a^+}\chi_1(r)=\lim_{r\rightarrow a^-}\chi_2(r). \label{boundary1}
\end{align}
Turning to Maxwell's equation, we see that Eq.~\Eqref{PDE} with $J^\phi$ given by \Eqref{J_def} reduces to
\begin{align}
 \sin\theta\sbrac{\frac{\dif}{\dif r}\brac{f\frac{\dif}{\dif r}\brac{r^2\chi}}-2\chi}=4\pi I\sin\theta\delta(\cos\theta) f(r)^{1/2}\delta(r-a).
\end{align}
We integrate both sides of this equation over all $\theta$, followed by an integration over $a-\epsilon\leq r\leq a+\epsilon$. In the limit $\epsilon\rightarrow0$, we have 
\begin{align}
 \left.\frac{\dif}{\dif r}\brac{r^2\chi_1(r)} \right|_{r=a}-\left.\frac{\dif}{\dif r}\brac{r^2\chi_2(r)} \right|_{r=a}=2\pi I f(a)^{-1/2}.\label{boundary2}
\end{align}
The desired regular solution is obtained as follows: For $F^2$ to be regular as $r\rightarrow0$, we take $K=-V(0)$. This choice removes the divergent term at the origin, giving 
\begin{align}
 F^2\sim\frac{C^2\brac{1+\cos^2\theta}}{18m^2\ell^4}+\frac{C^2\brac{7+6\cos^2\theta}}{90m^2\ell^6}r^2+\mathcal{O}(r^4),
\end{align}
which has a finite limit as $r\rightarrow0$. Subsequently, Eqs.~\Eqref{boundary1} and \Eqref{boundary2} fixes $C$ and $I$ in terms of $B$. In this way, we have a regular solution for any small $a$, as long as it is non-zero. As an explicit numerical example, for a Hayward spacetime with parameters $m=1$ and $\ell=0.8$ with loop radius $a=0.5$, we have $K=-33.718$, $C=57.293B$, and $I=2.9066B$. (Up to five significant figures.) 

This introduction of a current loop marks a difference from the weakly magnetised Schwarzschild case, where solving Maxwell's equation results in a homogeneous magnetic field with no current sources considered in the domain of the solution. The magnetic field in the Schwarzschild case can be assumed to be sourced by currents outside the domain of consideration (large $r$), possibly created by charged flow of accretion disks. In the Hayward case, just solving the vacuum equation and assuming the source to be unspecified outside may only explain the near-uniform magnetic field at $r>r_c$ similar to the Schwarzschild case. At $r<r_c$, the magnetic field lines start to take a dipole-like loop structure whose field strength is singular at the origin. 
Just as in multipole solutions of non-relativistic magneticstatics, this indicates that there should exist a current source $J^\mu$ to support this structure of the magnetic field. We take the physical interpretation of $J^\mu$ being an idealised model for the `accretion disk' sourcing the magnetic field. If $a$ is taken to be small up to the infinitesimal limit, this may serve as a simple toy model used to regularise the singularity in place of, perhaps, a more complete quantum field theory treatment of the electromagnetic field.  

For the remainder of this paper, we shall assume $r>a$ for all regions of interest. Hence we have
\begin{align}
 A=r^2\chi_1(r)\sin^2\theta\,\dif\phi=\frac{B}{2}\brac{1-\frac{4m\ell^2}{r^3}}r^2\sin^2\theta\,\dif\phi, \quad r>a,
\end{align}
as desired. The non-zero components of the Maxwell tensor are
\begin{align}
 F_{r\phi}=-F_{\phi r}=\brac{\chi_1 r^2}'\sin^2\theta,\quad F_{\theta\phi}=-F_{\phi\theta}=2\chi_1 r^2\sin\theta\cos\theta,
\end{align}
where primes denote derivatives with respect to $r$. In the case $\ell=0$, the solution reduces to the test magnetic field around the Schwarzschild black hole which was obtained by Wald in \cite{Wald:1974np}.

To compute the explicit form of the magnetic field, let $\xi^\mu=\frac{1}{\sqrt{f}}\delta_t^\mu$ be the unit time-like Killing vector of the spacetime. The covariant components of the magnetic field are given by
\begin{align}
 B_\mu=-\half\epsilon_{\mu\nu\rho\sigma}\xi^\nu F^{\rho\sigma},
\end{align}
where $\epsilon_{\mu\nu\rho\sigma}$ is the anti-symmetric Levi--Civita form with $\epsilon_{tr\theta\phi}=+\sqrt{|\det g|}=r^2\sin\theta$. The vector field representing the magnetic field is $\vec{B}=B^\mu\partial_\mu$. Explicitly, 
\begin{align}
 \vec{B}=f^{1/2}\sbrac{2\chi_1\cos\theta\;\vec{e}_r-\frac{(\chi_1 r^2)'}{r^2}\sin\theta\;\vec{e}_\theta}, \label{B_field_eqn}
\end{align}
where we have used the notation $\vec{e}_r=\partial_r$ and $\vec{e}_\theta=\partial_\theta$ for the basis vectors.

To visualise the magnetic field lines, we transform to cylindrical coordinates by $\rho=r\sin\theta$ and $z=r\cos\theta$. In these coordinates, the magnetic field is 
\begin{align}
 \vec{B}=f^{1/2}\sbrac{-\chi_1'r\sin\theta\cos\theta\;\vec{e}_\rho+\brac{2\chi+\chi_1' r\sin^2\theta}\;\vec{e}_z}. \label{B_chi1}
\end{align}
The presence of $\chi$ which is non-constant for $\ell\neq0$ makes the magnetic field non-uniform and has nontrivial $\rho$ and $z$ components. Nevertheless, as $r\rightarrow\infty$, the magnetic field asymptotically approaches a uniform configuration pointing along the $z$-direction. Only when $\ell=0$, the whole magnetic field is uniform along $z$ throughout the spacetime, recovering the weakly-magnetised Schwarzschild solution \cite{Wald:1974np}. 

\begin{figure}
 \centering 
 \begin{subfigure}[b]{0.49\textwidth}
    \centering
    \includegraphics[width=0.85\textwidth]{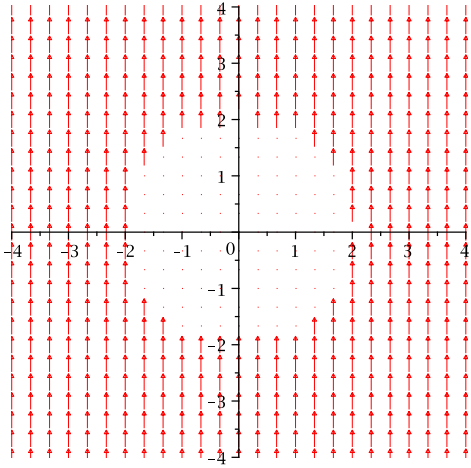}
    \caption{$\ell=0$ (Schwarzchild).}
    \label{fig_Bfield_a}
  \end{subfigure}
  \begin{subfigure}[b]{0.49\textwidth}
    \centering
    \includegraphics[width=0.85\textwidth]{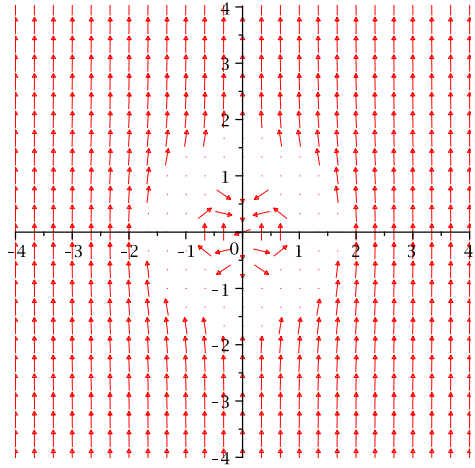}
    \caption{$\ell=0.6$.}
    \label{fig_Bfield_b}
  \end{subfigure}
  \begin{subfigure}[b]{0.49\textwidth}
    \centering
    \includegraphics[width=0.85\textwidth]{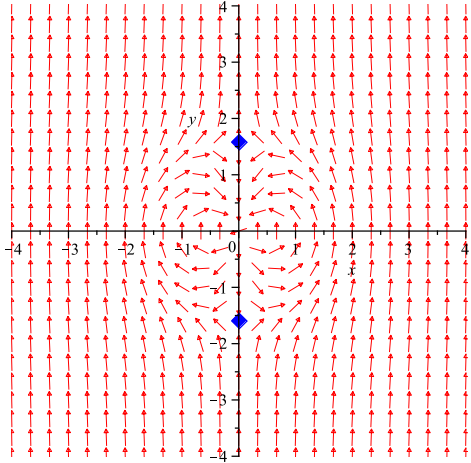}
    \caption{$\ell=1$.}
    \label{fig_Bfield_c}
  \end{subfigure}
  \begin{subfigure}[b]{0.49\textwidth}
    \centering
    \includegraphics[width=0.85\textwidth]{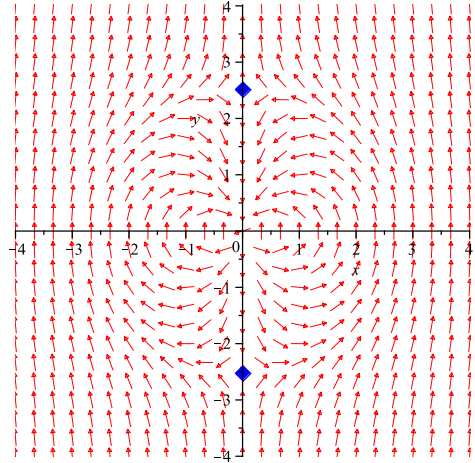}
    \caption{$\ell=2$.}
    \label{fig_Bfield_d}
  \end{subfigure}
  \caption{Plots of vector field $\vec{B}$ on the Hayward spacetime, for $m=1$ in cylindrical coordinates of fixed azimuth $\phi$. The empty regions are where $f<0$ and the magnetic field is undefined. In Figs.~\ref{fig_Bfield_c} and \ref{fig_Bfield_d}, the solid blue diamonds indicate points where $\vec{B}=0$. Here, the current loop radius $a$ is taken to be very small compared to the scales of the figure, and the vectors depicted correspond to the $\chi_1$ solution.}
  \label{fig_Bfield}
\end{figure}

The plots of the magnetic field are given in Fig.~\ref{fig_Bfield}. We take $a$ to be small such that the vectors depicted correspond to the exterior $r>a$ solution given by Eq.~\Eqref{B_chi1}. The empty regions correspond to the non-static region with $f<0$, for which $\vec{B}$ is undefined. Hence the plot for $\ell=0$ in Fig.~\ref{fig_Bfield_a} has an empty disk of radius $2m$ corresponding to the region interior to the Schwarzschild black hole. Outside the horizon, we see the magnetic field pointing uniformly in the $z$-direction, in accordance to the pure Schwarzschild configuration \cite{Wald:1974np}. The magnetic field for the Hayward black hole is shown in Fig.~\ref{fig_Bfield_b}, plotted for the case $m=1$ and $\ell=0.6$. The region exterior to the outer horizon shows some deviation from being perfectly uniform along the $z$-axis, and the static region inside the inner horizon shows looped magnetic field lines, reminiscent to the magnetic field lines emanating from an isolated point dipole at the origin. 

Figures \ref{fig_Bfield_c} and \ref{fig_Bfield_d} show the horizonless spacetime for $\ell=1$ and $\ell=2$, respectively. In these cases, we see the loop configuration of the magnetic field lines transitioning continuously to a $z$-uniform configuration as the distance from the origin is increased. As the Hayward parameter $\ell$ is increased, the dipole-like configuration takes up a larger space. 

The transition between dipole loops to the uniform parallel field lines can be understood through the characteristic radius $r_c$ defined in Eq.~\Eqref{rc_def}. In particular, $r_c$ is also a root of $\chi(r)$. This root gives two regimes for the potential:
\begin{align*}
 \chi(r)\sim\left\{\begin{array}{cc}
                    -\frac{1}{2}B_0\frac{4m\ell^2}{r^3}, & a<r\ll r_c,\\
                    \frac{1}{2}B_0, & r\gg r_c.
                   \end{array}\right.
\end{align*}
%

If the spacetime is the Hayward black hole, the characteristic radius always lies inside the non-extremal black hole horizon. Otherwise, $r=r_c=\brac{4m\ell^2}^{1/3}$ is precisely on the extremal horizon itself. This can be seen\footnote{Alternatively, we can also see this by noting that $r_c$ is a local minimum of $f(r)$ in the domain $r>0$, and must lie between the two roots of $f(r)$.} by checking 
\begin{align}
 f(r_c)=1-\frac{4^{2/3}}{3}\brac{\frac{m}{\ell}}^{2/3}.
\end{align}
For the black hole case, we have $\frac{m^2}{\ell^2}\geq\frac{27}{16}$, or $\brac{\frac{m}{\ell}}^{2/3}\geq\frac{3}{4^{1/3}}$. Therefore, we have 
\begin{align}
 f(r_c)\leq0\quad\mbox{ if }\quad \frac{m^2}{\ell^2}\geq\frac{27}{16}.
\end{align}
In other words, the magnetic field at the exterior of the Hayward black hole is always the uniform field regime. As we approach the horizon, $\vec{B}$ deviates away from being perfectly parallel. As can be seen in the example of Fig.~\ref{fig_Bfield_b}. The dipole loop field lines are strictly in the inner core. 

If $\frac{m^2}{\ell^2}<\frac{27}{16}$, no horizons are present. Therefore the entire spacetime is static and hence we see the interior dipole loop field lines at $r<r_c$ continuously deform into the uniform parallel lines at $r>r_c$, as can be seen in Figs.~\ref{fig_Bfield_c} and \ref{fig_Bfield_d}. Note that at the characteristic radius $r=r_c$, the magnetic field is precisely zero on the $z$-axis. In fact, we will show in Sec.~\ref{sec_NonEquatorial} charged particles can exist in stable equilibrium on these two points.

\section{Equations of motion for a charged particle the magnetised Hayward spacetime} \label{sec_EOM}

The motion of a time-like particle carrying charge per mass $e$ is described by a parametrised curve 
\begin{align*}
 x^\mu(\tau)=\brac{t(\tau),r(\tau),\theta(\tau),\phi(\tau)},
\end{align*}
where $\tau$ is the proper time of the particle. We will assume that the charged particle does not interact directly with the matter sources of the Hayward spacetime \Eqref{stress} apart from the gravity due to the Hayward geometry. The charged particle will have the usual Lorentz interaction with the test magnetic field derived in the previous section. Therefore the motion is governed by the Lagrangian $\mathcal{L}=\half g_{\mu\nu}\dot{x}^\mu\dot{x}^\nu+eA_\mu\dot{x}^\mu$, where over-dots denote derivatives with respect to $\tau$. For a weakly-magnetised Hayward spacetime, the Lagrangian is explicitly
\begin{align}
 \mathcal{L}=\half\brac{-f\dot{t}^2+\frac{\dot{r}^2}{f}+r^2\dot{\theta}^2+r^2\sin^2\theta\,\dot{\phi}^2}+e\chi_1 r^2\sin^2\theta\,\dot{\phi}, 
\end{align}
The conserved quantities are energy $E$ and angular momentum $L$ such that
\begin{align}
 \dot{t}=\frac{E}{f},\quad \dot{\phi}=\frac{L}{r^2\sin^2\theta}-e\chi_1.\label{tdotphidot}
\end{align}
The equations of motion in the $r$- and $\theta$-directions can be derived from the Euler--Lagrange equation, giving 
\begin{align}
 \ddot{r}&=\frac{f'}{2f}\dot{r}^2+rf\dot{\theta}^2-\frac{f'E^2}{2f}+f\brac{\frac{L^2}{r^3\sin^2\theta}-e^2\chi_1^2r\sin^2\theta}+fe\chi_1'\brac{L-e\chi_1 r^2\sin^2\theta},\label{ddot_r}\\
 \ddot{\theta}&=-\frac{2}{r}\dot{r}\dot{\theta}+\frac{L^2\cos\theta}{r^4\sin^3\theta}-e^2\chi_1^2\sin\theta\cos\theta.\label{ddot_theta}
\end{align}
As before, primes denote derivatives with respect to $r$. The time-like character of the particle gives a first integral equation $g_{\mu\nu}\dot{x}^\mu\dot{x}^\nu=\epsilon$, where $\epsilon$ is a negative constant. Eliminating $\dot{t}$ and $\dot{\phi}$ in favour of $E$ and $L$, this leads to 
\begin{align}
 \frac{\dot{r}^2}{f}+r^2\dot{\theta}^2-\frac{E^2}{f}+r^2\sin^2\theta\brac{\frac{L}{r^2\sin^2\theta}-e\chi_1}^2=\epsilon.\label{FirstIntegral}
\end{align}
We assume the proper time parameter is appropriately rescaled such that $\epsilon=-1$. Upon rearranging, Eq.~\Eqref{FirstIntegral} can be expressed in effective potential form, 
\begin{align}
 \dot{r}^2+r^2f\dot{\theta}^2=\mathcal{E}-\mathcal{U}, \label{eff_energy_eqn}
\end{align}
where $\mathcal{E}=E^2$ and the \emph{effective potential} is given by 
\begin{align}
 \mathcal{U}(r,\theta)=f\sbrac{r^2\sin^2\theta\brac{\frac{L}{r^2\sin^2\theta}-e\chi_1}^2+1}
\end{align}
From Eq.~\Eqref{eff_energy_eqn}, we see that the particles must be confined to move in a domain where $\mathcal{E}-\mathcal{U}\geq0$. 

Since $e$ and $B_0/2$ always appear together as products, it will be convenient to define $\beta=\half eB_0$. Therefore the motion of our charged paraticle will be parametrised by $\beta$, $E$ and $L$. The  spacetime geometry itself is parametrised by the mass parameter $m$ and Hayward parameter $\ell$, as usual.  The equations of motion are generally non-separable, except for equatorial orbits. Therefore an analytical study of these equations are possible only in specific cases and perturbations thereof. For a general situation, a trajectory is obtained by solving Eqs.~\Eqref{ddot_r} and \Eqref{ddot_theta} numerically using the fourth-order Runge--Kutta algorithm implemented in C. The first integral \Eqref{FirstIntegral} is used as a consistency check throughout the numerical solution.

\section{Equatorial motion} \label{sec_Equatorial}

It can be shown that $\theta=\frac{\pi}{2}=\mathrm{constant}$ is a solution to the $\theta$-equation \Eqref{ddot_theta}. Therefore in such a case the motion is confined to the equatorial plane. The effective potential is now a function of $r$ only, and its structure can be easily studied in one dimensional graphs of $r$ against $\mathcal{U}$. In the following we will also study its circular orbits as well as the curly behaviour of the trajectories. 

\subsection{Effective potential}

\begin{figure}
 \centering 
 \includegraphics[scale=0.74]{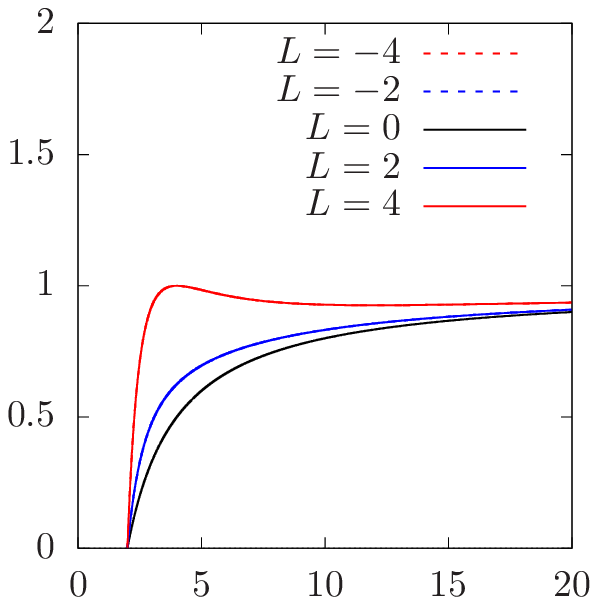}
 \includegraphics[scale=0.74]{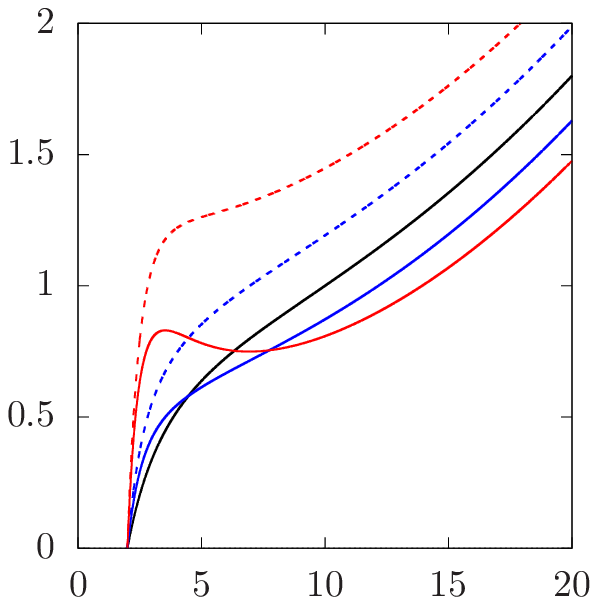}
 \includegraphics[scale=0.74]{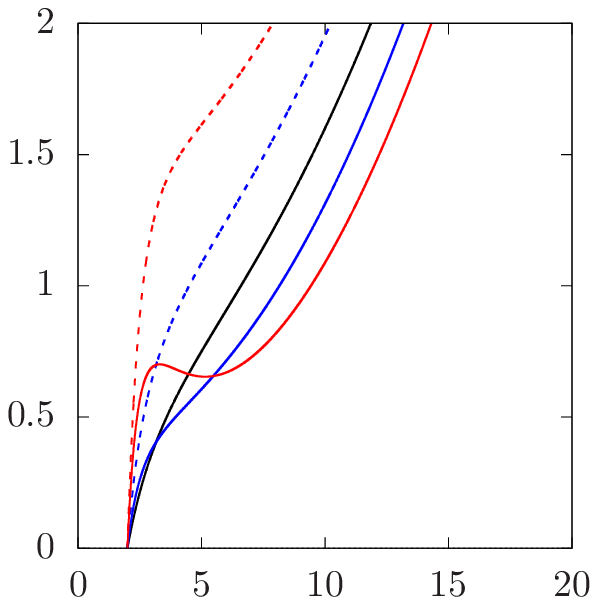}
 
 \includegraphics[scale=0.74]{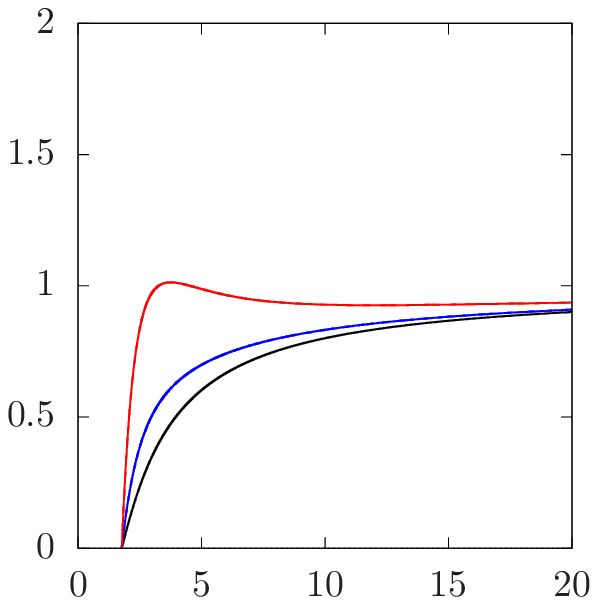}
 \includegraphics[scale=0.74]{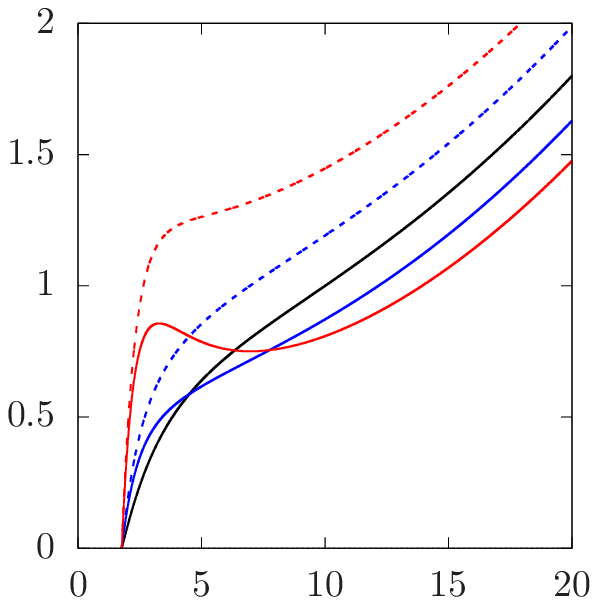}
 \includegraphics[scale=0.74]{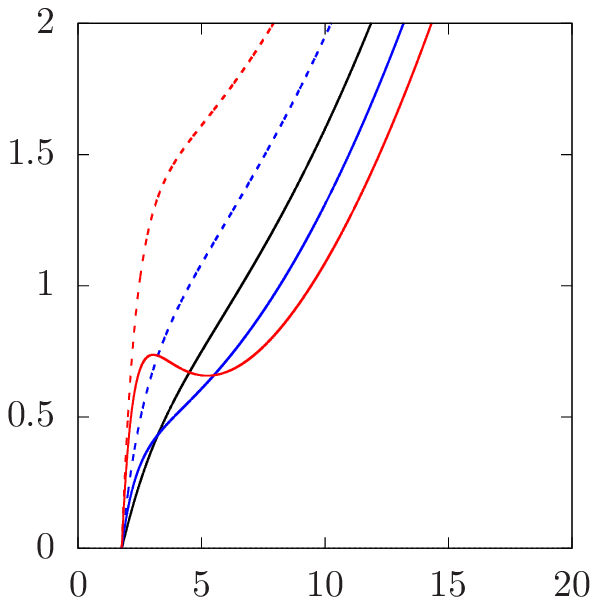}
 
 \includegraphics[scale=0.74]{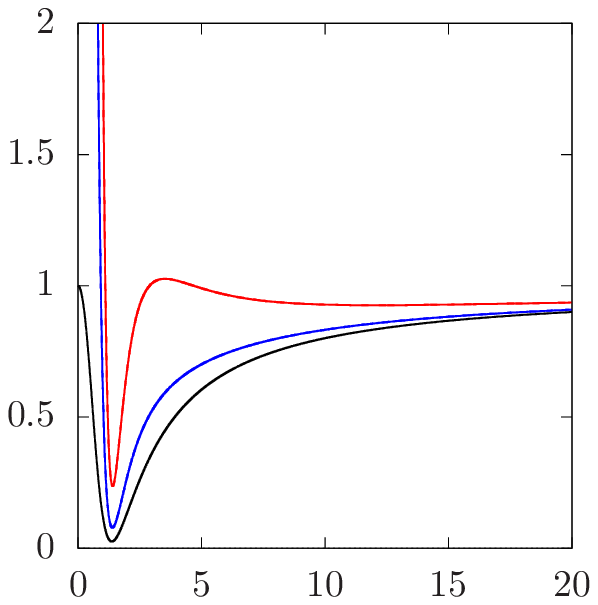}
 \includegraphics[scale=0.74]{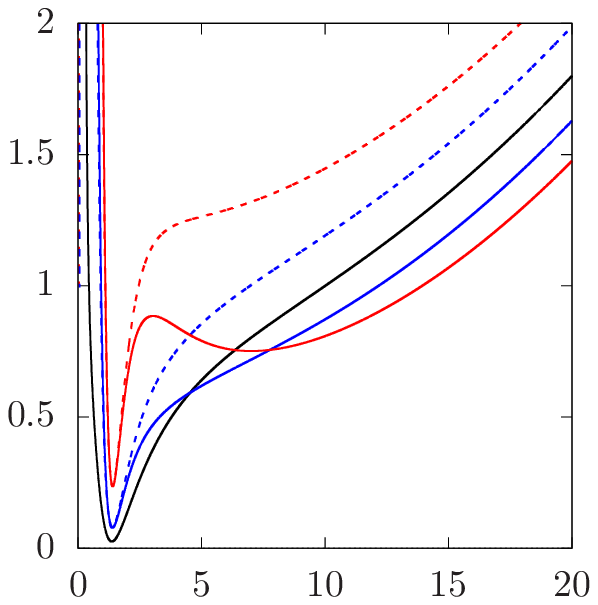}
 \includegraphics[scale=0.74]{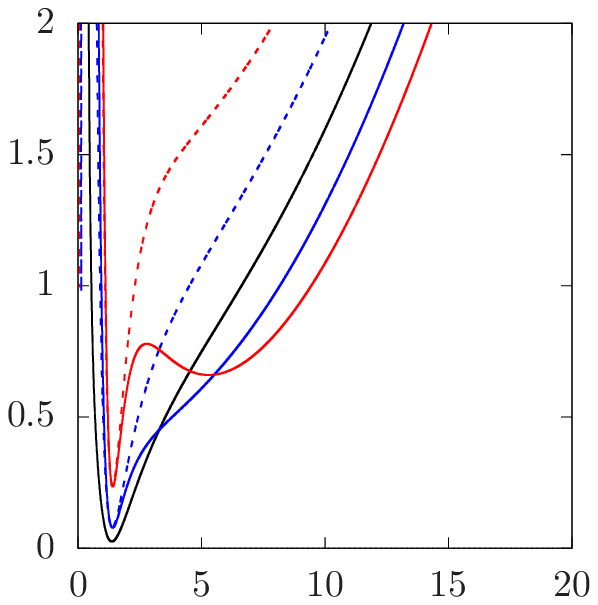}
  \caption{Plots of $r$ vs $\mathcal{U}$ for $m=1$. The top, middle, and bottom rows correspond to $\ell=0$, $0.6$, and $0.8$, respectively. The left, middle, and right columns correspond to $\beta=0$, $0.05$, and $0.1$, respectively.}
  \label{fig_EqPot}
\end{figure}

By numerical exploration, we find that for particles around the Hayward black hole $\frac{\ell}{m}<\frac{4}{3\sqrt{3}}$, the effective potential tend to have one local minima and one local maxima. The potential terminates at the horizon $r=r_+$ with $\mathcal{U}=0$. For $\beta\neq0$ we have $\lim_{r\rightarrow\infty}\mathcal{U}=\infty$. All charged particles cannot escape to infinity. Effective potentials around the Hayward black hole are plotted in the middle column of Fig.~\ref{fig_EqPot}, particularly $\ell=0.6$. In each of these plots, we see that as $L$ is adjusted, the local minimum and maximum coalesce into a point $\mathcal{U}''=0$, giving the innermost stable circular orbit (ISCO). 

For spacetimes with $\frac{\ell}{m}>\frac{4}{3\sqrt{3}}$ we have no event horizon. As in the black hole case, we have $\lim_{r\rightarrow \infty}\mathcal{U}=\infty$. Additionally, we now also have $\lim_{r\rightarrow0}\mathcal{U}=\infty$ for $L\neq0$, indicating an infinite potential barrier preventing particles of non-zero angular momentum from reaching the origin. For certain ranges of $L$, the effective potentials tend to have a double-well structure, such as the ones shown in the middle pane of the bottom row of Fig.~\ref{fig_EqPot}. This means that we have two disconnected domains of bound orbits for the same set of parameters, as well as having three circular orbits; two stable and one unstable. As $L$ is adjusted, two of the maxima may coalesce and subsequently vanish, leaving behind one minima. Since $\mathcal{U}$ goes to infinity both for $r\rightarrow0$ and $r\rightarrow\infty$, there is always at least one local minima of the potential.

\subsection{Curly orbits} \label{subsec_curly}

A characteristic influence of a magnetic field on a charged particle is the tendency of the trajectory to `curl up'. More precisely, from Eq.~\Eqref{tdotphidot} for $\dot{\phi}$, we see that $\dot{\phi}$ changes sign when $\frac{L}{r^2}=e\chi$, or, at the roots of
\begin{align}
 \beta r^3-Lr-4m\ell^2\beta=0.
\end{align}
Assuming for the moment that there exist one positive root, which we denote by $r_*$. This radial coordinate indicates the point where the angular motion changes direction. (That is, $\dot{\phi}$ changes sign.) We denote by $\mathcal{U}_*=\left.\mathcal{U}\right|_{r=r_*}$ the value of the effective potential at this radius. The possible orbits may have different appearances depending on the particle's energy relative to $\mathcal{U}_*$, and is completely analogous to the kinds of orbits already discussed in other magnetised spacetimes, such as \cite{Frolov:2010mi,Kolos:2015iva,Lim:2020fnx}. Let us now discuss these orbits in the present context of the magnetised Hayward spacetimes.

\begin{figure}
 \centering 
 \begin{subfigure}[b]{0.49\textwidth}
    \centering
    \includegraphics[width=\textwidth]{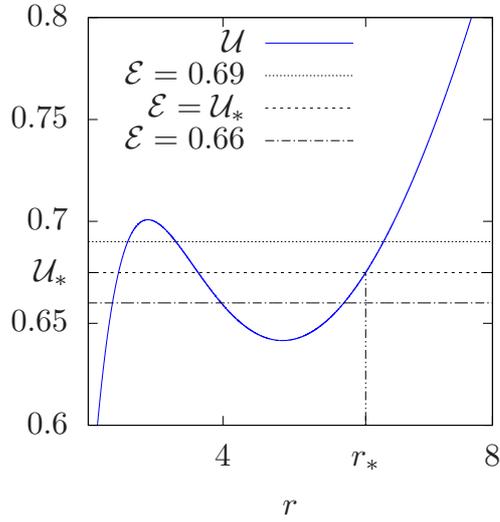}
    \caption{Effective potential.}
    \label{fig_CurlyEx_pot}
  \end{subfigure}
  \begin{subfigure}[b]{0.49\textwidth}
    \centering
    \includegraphics[width=\textwidth]{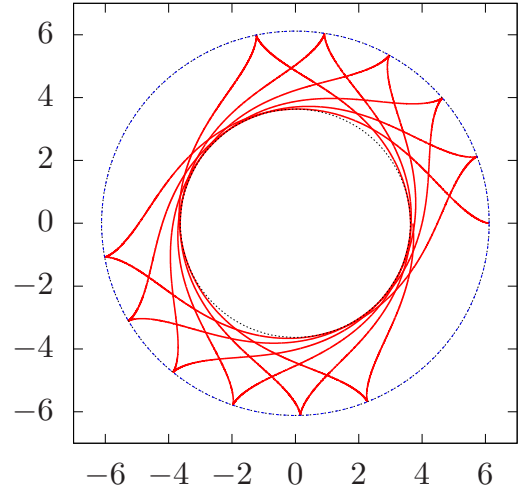}
    \caption{$\mathcal{E}=\mathcal{U}_*$.}
    \label{fig_CurlyEx_cusp}
  \end{subfigure}
  \begin{subfigure}[b]{0.49\textwidth}
    \centering
    \includegraphics[width=\textwidth]{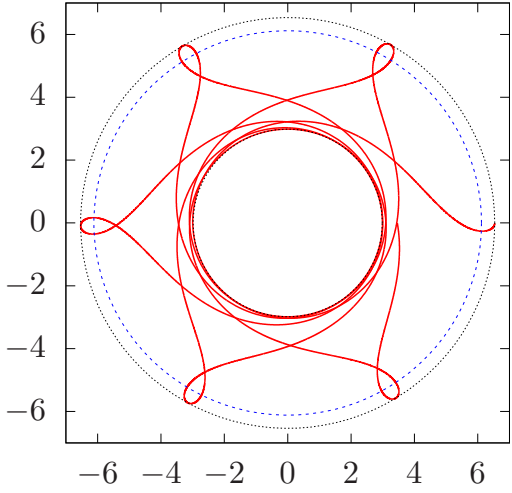}
    \caption{$\mathcal{E}=0.69(>\mathcal{U}_*)$.}
    \label{fig_CurlyEx_curl}
  \end{subfigure}
  \begin{subfigure}[b]{0.49\textwidth}
    \centering
    \includegraphics[width=\textwidth]{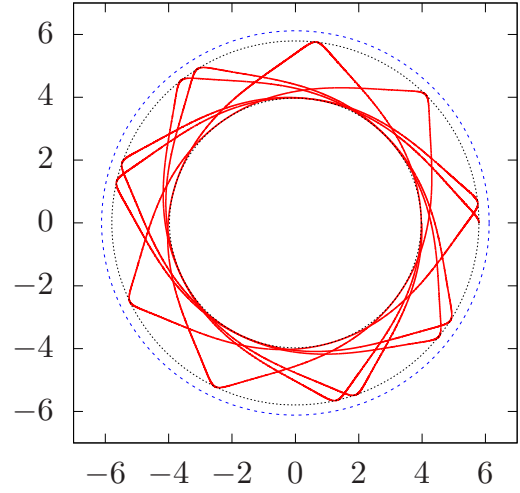}
    \caption{$\mathcal{E}=0.66(<\mathcal{U}_*)$.}
    \label{fig_CurlyEx_reg}
  \end{subfigure}
  \caption{Examples of orbits of a particle with charge parameter $\beta=0.1$ and angular momentum $L=3.7$ in a Hayward spacetime of $m=1$ and $\ell=0.8$. For these parameters, $\dot{\phi}=0$ at $r_*\simeq 6.117$, and $\mathcal{U}(r_*)=\mathcal{U}_*\simeq0.6749$. The effective potential is shown in Fig.~\ref{fig_CurlyEx_pot}. Orbits with $\mathcal{E}=\mathcal{U}_*$ has sharp cusps, as shown in Fig.~\ref{fig_CurlyEx_cusp}. An example of an curly orbit where $\mathcal{E}>\mathcal{U}_*$ is shown in Fig.~\ref{fig_CurlyEx_curl}, with $\mathcal{E}=0.7$. Figure \ref{fig_CurlyEx_reg} shows a regular orbit with $\mathcal{E}=0.66<\mathcal{U}_*$, where $\dot{\phi}$ never reaches zero.}
  \label{fig_CurlyEx}
\end{figure}

In the following, we shall demonstrate the appearance of such orbits using a concrete example for $m=1$, $\ell=0.8$, and $\beta=0.1$. Its corresponding effective potential is plotted in Fig.~\ref{fig_CurlyEx_pot}. If the particle has energy such that $\mathcal{E}=\mathcal{U}_*$, the turning point of $r$ coincides with the turning point of $\phi$. In this situation the trajectory forms sharp cusps, such as in Fig.~\ref{fig_CurlyEx_cusp}. If the particle has energy $\mathcal{E}>\mathcal{U}_*$, then the particle crosses $r_*$ at various occasions throughout its motion, meaning that the direction of $\dot{\phi}$ changes. This leads to the shape of the curled trajectories, where an example is shown in Fig.~\ref{fig_CurlyEx_curl}. On the other hand, if $\mathcal{E}<\mathcal{U}_*$, the point $r=r_*$ is not accessible by the particle, and $\dot{\phi}$ does not change sign throughout its motion. An example of this is shown in Fig.~\ref{fig_CurlyEx_reg}. 

\subsection{Circular orbits} 

Circular orbits are found from the condition
\begin{align}
 \mathcal{U}&=\mathcal{E},\quad
\mathcal{U}'=0.\label{discr}
\end{align}
the second equation of \Eqref{discr} can be solved for $L$ to give the corresponding angular momenta for the circular orbits. For circular orbits of radius $r=r_0$, this gives two sets of solutions,
\begin{align}
 L_\pm&=\frac{1}{\brac{r_0^6-3mr_0^5+4m\ell^2 r_0^3+4m^2\ell^4}r_0}\nonumber\\
   &\quad\times\Big[-m\beta\brac{r_0^8+6\ell^2 r_0^6+24m\ell^4r_0^3-20m\ell^2 r_0^5+24m^2\ell^6-8m^2\ell^4r_0^2}\pm\sqrt{\Delta}\Big], \label{ISCO_L_eqn}
\end{align}
where 
\begin{align}
 \Delta=\brac{r_0^3+2m\ell^2-2mr_0^2}^2\brac{r_0^3+2m\ell^2}^4\beta^2-mr_0^6\brac{4m\ell^2-r_0^3}\brac{4m\ell^2r_0^3+4m^2\ell^4+r_0^6-3mr_0^5}. \label{Delta_def}
\end{align}
The corresponding energy $E=\sqrt{\mathcal{E}}$ is obtained from the first equation of \Eqref{discr} 
\begin{align}
 E_\pm=\frac{1}{r_0^2}\sqrt{\frac{\brac{r_0^3+2m\ell^2-2mr_0^2}\sbrac{r_0^2(r_0^2+L_\pm^2)+2L_\pm\beta (4m\ell^2-r_0^3)r+(4m\ell^2-r_0^3)^2\beta^2}}{\brac{r_0^3+2m\ell^2}}}. \label{E_circ}
\end{align}

For each set of circular orbits $(L_\pm, E_\pm)$, the stability of the circular orbits can be determined by checking the sign of $\mathcal{U}''$. For Hayward black holes, there exist an \emph{innermost stable circular orbit} (ISCO) whose radius we denote by $r_{\mathrm{ISCO}\pm}$. For $r<r_{\mathrm{ISCO}\pm}$ we see that $\mathcal{U}''<0$ where the orbits are unstable. Conversely, for larger radii $r>r_{\mathrm{ISCO}}$ the circular orbits are stable with $\mathcal{U}''>0$.

\begin{figure}
 \centering 
 \begin{subfigure}[b]{0.49\textwidth}
    \centering
    \includegraphics[width=\textwidth]{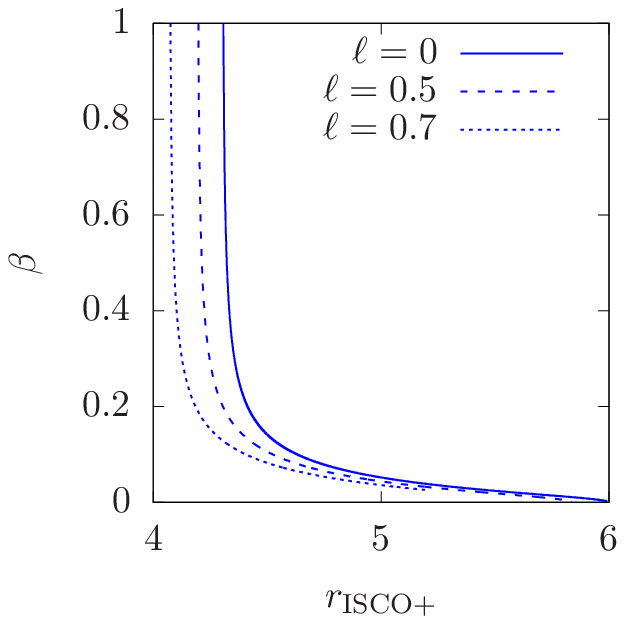}
    \caption{ISCOs with $L=L_+$.}
    \label{fig_ISCO_L1}
  \end{subfigure}
  \begin{subfigure}[b]{0.49\textwidth}
    \centering
    \includegraphics[width=\textwidth]{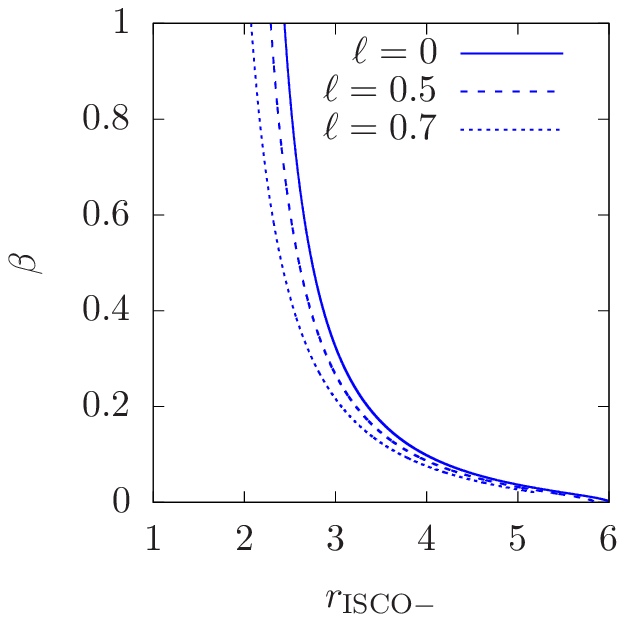}
    \caption{ISCOs with $L=L_-$.}
    \label{fig_ISCO_L2}
  \end{subfigure}
  \caption{Plots of ISCO radii vs $\beta$ around a Hayward black hole for the two branches of angular momenta $L=L_\pm$ given by Eq.~\Eqref{ISCO_L_eqn}.}
  \label{fig_ISCO}
\end{figure}

In the magnetised Schwarzschild case, it was found that the presence of Lorentz interaction between the particle and the magnetic field allows the ISCO to have a smaller radius than that of the pure Schwarzschild case of $r=6m$ \cite{Frolov:2010mi}. This fact is potentially significant as it may have observational consequences for astrophysical black holes \cite{Frolov:2014zia}. These are reproduced in the solid curves of Fig.~\ref{fig_ISCO} where $\ell=0$. Furthermore, for the same $\beta$, the inclusion of the Hayward parameter $\ell>0$ further reduces the radii of the ISCOs, as shown in the dashed and dotted curves of Fig.~\ref{fig_ISCO}. 

\begin{figure}
 \centering 
 \includegraphics{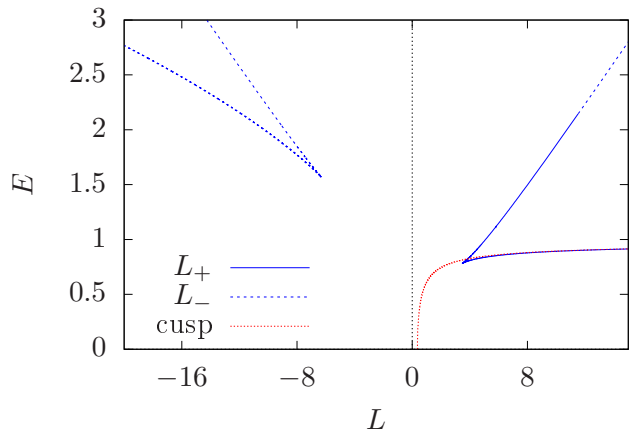}
 \caption{Parameter space for a charged particle with charged parameter $\beta=0.1$ around a Hayward black hole of $m=1$, $\ell=0.3$.}
 \label{fig_ParamSpace_BH}
\end{figure}

The various kinds of equatorial orbits can be summarised in an $(L,E)$-parameter space, such as in Fig.~\ref{fig_ParamSpace_BH}. In that figure, the blue curves are the values of $L$ and $E$ corresponding to circular orbits. This is essentially obtained by plotting Eqs.~\Eqref{ISCO_L_eqn} and \Eqref{E_circ} as a parametric curve in $r_0$. The red curve are the parameters corresponding to orbits with cusps. Therefore if these orbits are bound, those with energies above this curve are curly orbits, and those below it are orbits without curls.

We notice that circular orbits for $(L_+,E_+)$ consists of two branches connecting at a sharp cusp. In the example shown in Fig.~\ref{fig_ParamSpace_BH}, this is the solid blue curve. The lower branch corresponds to the stable circular orbits of radii $r_{\mathrm{ISCO}+}<r_0<\infty$. After passing through the cusp, we are at the upper branch corresponding to the unstable circular orbits for $r_{\mathrm{ISCO}+}<r_0<r_\Delta$, where $r_\Delta$ is the root of the function $\Delta$ defined in Eq.~\Eqref{Delta_def}. For radii $r_0<r_\Delta$, $L_+$ becomes complex-valued and hence no circular orbits exist.

The other set of circular orbits are $(L_-,E_-)$ consists of two disconnected components, shown as the blue dotted curves in Fig.~\ref{fig_ParamSpace_BH}. The connected component on the left consists of two branches meeting at a cusp. The lower branch corresponds to stable circular orbits of radii $r_{\mathrm{ISCO}-}<r_0<\infty$, where the cusp is the ISCO. The upper branch is the unstable circular orbits of radii $r_{\mathrm{ISCO}-}<r_0<r_\infty$, where $r_\infty$ is the value of $r_0$ where $L_-$ diverges. (That is, where the denominator in Eq.~\Eqref{ISCO_L_eqn} goes to zero.) The other component of the unstable circular orbit is $r_\infty<r_0<r_\Delta$ on the right, where $L_-$ is positive. It joins the $L_+$ unstable circular orbit at $r_0=r_\Delta$, for which $L_+=L_-$.

For the horizonless Hayward spacetimes of $\ell>\frac{4}{3\sqrt{3}}m$, the structure of the parameter space is slightly more complicated. This is mainly due to the fact that double potential wells are possible here. Hence there might be up to three points for which $\mathcal{U}=\mathcal{U}'=0$ corresponding to three circular orbits for a fixed $L$. Two of them are stable and one unstable. In this case there are no ISCOs here, But rather two distinct points for which $\mathcal{U}=\mathcal{U}'=\mathcal{U}''=0$, which we denote by 
\begin{align}
 r_{\mathrm{OCCO}\pm},\quad \mbox{ and }\quad r_{\mathrm{ICCO}\pm},
\end{align}
where `OCCO' and `ICCO' is what we shall call \emph{outer critical circular orbit} and \emph{inner critical circular orbit}, respectively. As such we have the association $r_{\mathrm{OCCO}\pm}>r_{\mathrm{ICCO}\pm}$. These correspond to the points when two circular orbits coalesce.

\begin{figure}
 \centering 
 \begin{subfigure}[b]{0.49\textwidth}
    \centering
    \includegraphics[width=\textwidth]{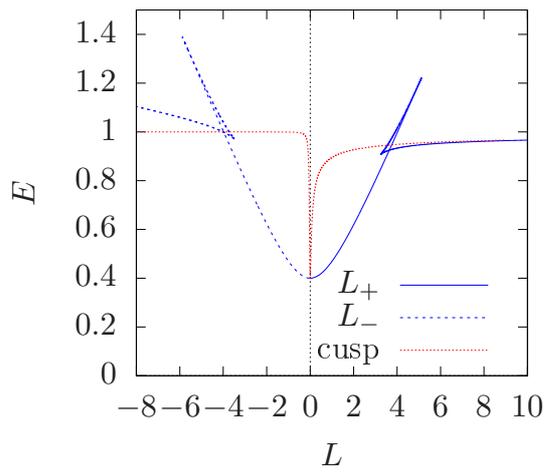}
    \caption{$\beta=0.01$.}
    \label{fig_ParamSpaceNoa}
  \end{subfigure}
  \begin{subfigure}[b]{0.49\textwidth}
    \centering
    \includegraphics[width=\textwidth]{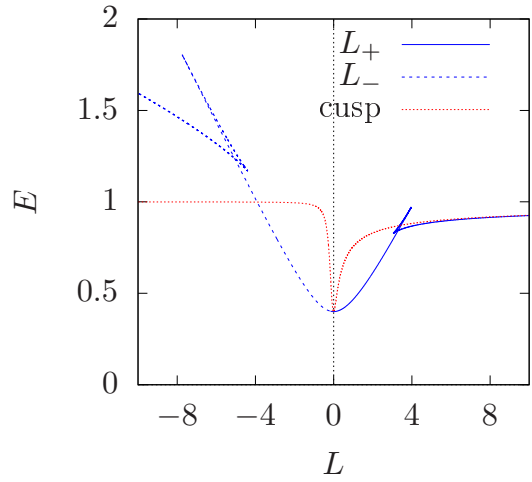}
    \caption{$\beta=0.05$.}
    \label{fig_ParamSpaceNob}
  \end{subfigure}
  \begin{subfigure}[b]{0.49\textwidth}
    \centering
    \includegraphics[width=\textwidth]{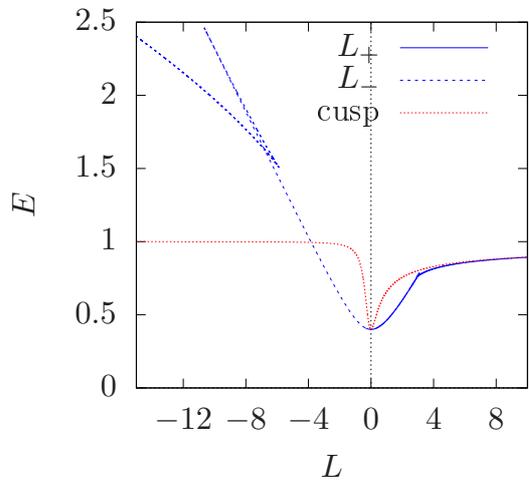}
    \caption{$\beta=0.1$.}
    \label{fig_ParamSpaceNoc}
  \end{subfigure}
  \begin{subfigure}[b]{0.49\textwidth}
    \centering
    \includegraphics[width=\textwidth]{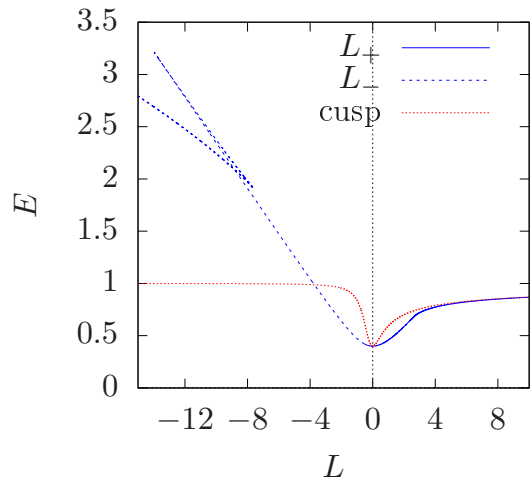}
    \caption{$\beta=0.15$.}
    \label{fig_ParamSpaceNod}
  \end{subfigure}
  \caption{Parameter space for a particle in a spacetime with $m=1$, $\ell=1$.}
  \label{fig_ParamSpace}
\end{figure}

Let us first explore the parameter space by taking the example $m=1$, $\ell=1$, and $\beta=0.01$, depicted in Fig.~\ref{fig_ParamSpaceNoa}. The solid blue curve are circular orbits of $L=L_+$ and $E=E_+$. The curve coming in from positive infinite $L$ and ending at the first cusp correspond to the stable circular orbits of $r_{\mathrm{OCCO}+}<r_0<\infty$. The endpoint of this branch is $r_0=r_{\mathrm{OCCO}+}$ where we encounter the first cusp of $L_+$. Then, the branch located between two cusps is the unstable circular orbit of radii $r_{\mathrm{ICCO}+}<r_0<r_{\mathrm{OCCO}+}$, after which we have the third branch corresponding to stable circular orbits of $r_{\mathrm{ICCO}+}<r_0<0$. The end of this branch is simply a particle of zero angular momentum sitting at the origin.

The other set of circular orbits with $L=L_-$ and $E=E_+$ are the dotted blue curve on the left side of Fig.~\ref{fig_ParamSpaceNoa}. The structure is similar to that of $(L_+,E_+)$. Namely the lower branch between $L\rightarrow\infty$ and the lower cusp is the stable circular orbit of radii $r_{\mathrm{OCCO}-}<r_0<\infty$. The branch between the two cusps are the unstable circular orbits of radii $r_{\mathrm{ICCO}-}<r_0<r_{\mathrm{OCCO}-}$, and finally the branch between the second cusp extending to $(L,E)=(0,0)$ is the stable inner circular orbits of $0<r_0<r_{\mathrm{ICCO}-}$.

Figs.~\ref{fig_ParamSpaceNob}, \ref{fig_ParamSpaceNoc}, and \ref{fig_ParamSpaceNod} shows the parameter space structure as $\beta$ is increased. Notably we see that as the particle's charge increases, the swallowtail structure of the $(L_+,E_+)$ curve shrinks and disappears, resulting in a smooth curve of stable circular orbits. This `phase transition' indicates that the unstable circular orbit disappears. 

\subsection{Perturbations of circular orbits} \label{subsec_circular_perturb}

In the previous subsection, we have found circular orbits where the radius and polar angles are $r=r_0$ and $\theta=\frac{\pi}{2}$, respectively. Now we consider small perturbations about these orbits. Specifically, starting with particles of energy and angular momenta $E_\pm$ and $L_\pm$ as given by Eqs.~\Eqref{E_circ} and \Eqref{ISCO_L_eqn} respectively, we perturb its position slightly away from $(r,\theta)=(r_0,\frac{\pi}{2})$ by writing 
\begin{align}
 r(\tau)&=r_0+\lambda r_1(\tau),\quad \theta(\tau)=\frac{\pi}{2}+\lambda\theta_1(\tau),\label{perturb}
\end{align}
where $\lambda$ is a small perturbation parameter such that terms of order $\mathcal{O}(\lambda^2)$ will be neglected. Substitution of Eq.~\Eqref{perturb} into Eqs.~\Eqref{ddot_r} and \Eqref{ddot_theta} gives, up to first order in $\lambda$,
\begin{align}
 \ddot{r}_1=-\omega_{r\pm}^2r_1,\quad \ddot{\theta}_1=-\omega_{\theta\pm}^2\theta_1,
\end{align}
where 
\begin{align}
 \omega_{r\pm}^2&=\frac{1}{\brac{r_0^3+2m\ell^2}\brac{4r_0^3m\ell^2+4m^2\ell^4-3r_0^5+r_0^6}^2r_0^6}\nonumber\\
  &\quad\times\big[-2r_0^6m\brac{8\ell^4m^3r_0^2-24\ell^4m^2r_0^3+8m^3\ell^6+r_0^9-r_0^8m-12\ell^2r_0^6m+20\ell^2r0^5m^2}E_\pm^2\nonumber\\
  &\quad\hspace{0.5cm}+r_0^2\brac{12m^2\ell^4-4\ell^2r_0^2m^2+12r0^3m\ell^2-8r_0^5m+3r_0^6}\brac{r_0^3+2m\ell^2-2mr_0^2}^2L_\pm^2\nonumber\\
  &\quad\hspace{0.5cm}+24\beta\ell^2mr_0\brac{8m^2\ell^4-4\ell^2r_0^2m^2+8r_0^3m\ell^2-5r_0^5m+2r_0^6}\brac{r_0^3+2m\ell^2-2mr_0^2}^2 L_\pm\nonumber\\
  &\quad\hspace{0.5cm}+\beta^2\big(640m^4\ell^8-384\ell^6m^4r0^2+608\ell^6m^3r0^3-384\ell^4m^3r0^5+132\ell^4m^2r_0^6\nonumber\\
  &\quad\hspace{2cm}+12r_0^8m^2\ell^2-4m\ell^2r_0^9+r_0^{12}\big)\brac{r_0^3+2m\ell^2-2mr_0^2}^2\big],\\
\omega_{\theta\pm}^2&=\frac{L_\pm^2 r_0^2-\beta^2\brac{r_0^3-4m\ell^2}^2}{r_0^6}.
\end{align}
In essence, the calculation of $\omega_{r\pm}^2$ reproduces the stability analysis from studying $\mathcal{U}''$ in the equatorial plane of the previous subsection. In particular, if $\omega_{r\pm}^2$ is positive, the particle oscillates radially about $r_0$, indicating a stable orbit. Conversely if $\omega_{r\pm}^2$ is negative, $r_1(\tau)$ increases exponentially away from $r_0$, showing that the original circular orbit is unstable.


Turning to the equation for $\phi$, substitution of Eq.~\Eqref{perturb} into $\dot{\phi}$ in Eq.~\Eqref{tdotphidot} gives, to linear order in $\lambda$, 
\begin{align}
 \dot{\phi}&=\Omega_\phi-\eta r_1\lambda,\nonumber\\
 \Omega_\phi&=\frac{L_\pm r_0+\beta\brac{4m\ell^2-r_0^3}}{r_0^3},\quad\eta=\frac{2}{r_0^4}\brac{Lr_0+6\beta m\ell^2}.
\end{align}
Here $\Omega_\phi$ would be the angular velocity in the $\phi$ direction for the unperturbed circular motion. 

Suppose we perturb about stable\footnote{Such that $\omega_{r\pm}$ and $\omega_{\theta\pm}$ are real.} circular orbits, a particular solution takes the form
\begin{subequations} \label{perturb_soln}
\begin{align}
 r(\tau)&=r_0+a\cos\brac{\omega_{r\pm}\tau}+\mathcal{O}(a^2),\label{perturb_r_soln}\\ \theta(\tau)&=\frac{\pi}{2}+b\cos\brac{\omega_{\theta\pm}\tau}+\mathcal{O}(b^2),\label{perturb_theta_soln}\\
 \phi(\tau)&=\phi_0+\Omega_\phi\tau-\frac{\eta a}{\omega_{r\pm}}\sin\brac{\omega_{r\pm}\tau}+\mathcal{O}(a^2), \label{perturb_phi_soln}
\end{align}
\end{subequations}
where $a$ and $b$ are of the order $\lambda$, and $\phi_0$ is an arbitrary choice for the initial condition of $\phi(\tau)$. With this solution let us investigate some cases that reveal the physics behind the forces experienced by the charged particle. Most of the behaviour are already present in the magnetised Schwarzschild case that were studied in \cite{Frolov:2014zia,Kolos:2015iva}. The introduction of the Hayward parameter $\ell$ changes the quantitative values of the orbital parameters.

\begin{figure}
 \begin{subfigure}[b]{\textwidth}
    \centering
    \includegraphics[scale=1]{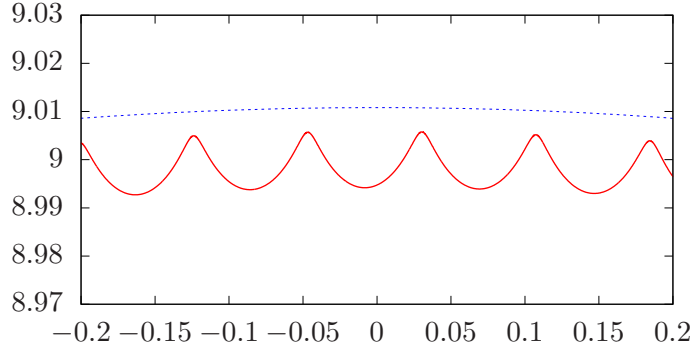}
    \caption{$a=r_*-r_0-0.005$, $\zeta=0.5387<1$ (curtate cycloid).}
    \label{fig_trochoid3}
  \end{subfigure}
 \begin{subfigure}[b]{\textwidth}
    \centering
    \includegraphics[scale=1]{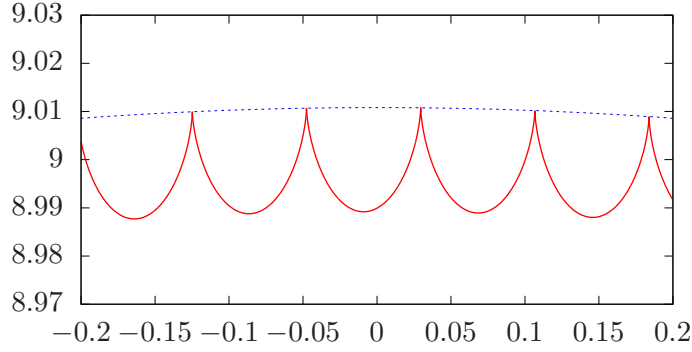}
    \caption{$a=r_*-r_0$, $\zeta=1.0018\approx1$ (common cycloid).}
    \label{fig_trochoid1}
  \end{subfigure}
  \begin{subfigure}[b]{\textwidth}
    \centering
    \includegraphics[scale=1]{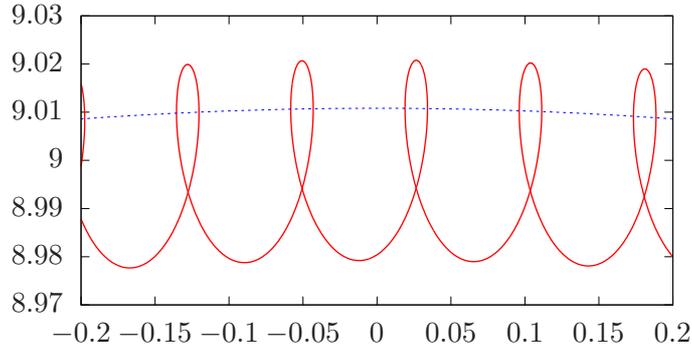}
    \caption{$a=r_*-r_0+0.01$, $\zeta=1.9280 > 1$ (prolate cycloid).}
    \label{fig_trochoid2}
  \end{subfigure}
  \caption{Perturbed circular orbits about $r_0=0.9$ of particles with charge parameter $\beta=0.6$ in a Hayward spacetime of $m=1$, $\ell=1$. The dotted blue line indicates the radius $r=r_*$, the point where $\dot{\phi}=0$. For these parameters $r_*=9.0108$. Numerical values reported here are shown up to 5 significant figures.}
  \label{fig_trochoid}
\end{figure}

\textbf{Cycloidal motion.} The `curly' orbits discussed in Sec.~\ref{subsec_curly} can be quantified in more detail in the case of perturbed circular orbit. More specifically Eq.~\Eqref{perturb_r_soln} and \Eqref{perturb_phi_soln} approximately describe the parametric equations for a \emph{trochoid} \cite{Frolov:2014zia}. The value of 
\begin{align}
 \zeta=\frac{\eta a}{\Omega_\phi}, \label{zeta_def}
\end{align}
determines the specific type of trochoid. In particular, the case $\zeta<1$ describes a \emph{curtate cycloid}, and is a case of $\mathcal{E}<\mathcal{U}_*$ where $\dot{\phi}$ does not change sign throughout the motion, as described in Sec.~\ref{subsec_curly}. If $\zeta=1$, we have the \emph{common cycloid} which is a particular case of the orbits with cusps $\mathcal{E}=\mathcal{U}_*$. If $\zeta>1$, we have the \emph{prolate cycloid}, which is an instance of the `curly orbits' of $\mathcal{E}>\mathcal{U}_*$ described in Sec.~\ref{subsec_curly}. Examples of these orbits are shown in Fig.~\ref{fig_trochoid}, for particles of $\beta=0.6$ perturbed around a circular orbit of $r_0=9$ in a spacetime with $m=1$ and $\ell=1$.

\begin{figure}
 \centering 
 \includegraphics{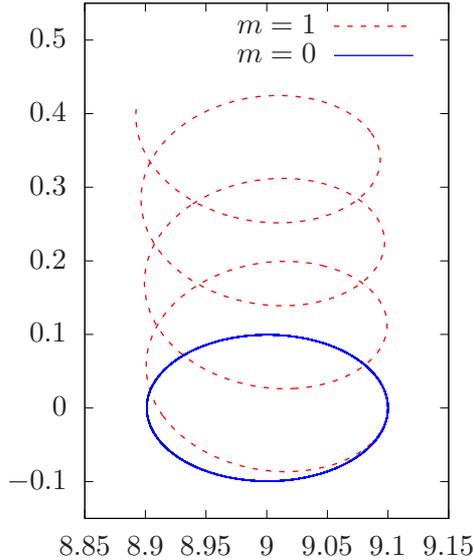}
 \caption{Orbits for $\beta=0.5$, $r_0=9$, $a=0.1$. The solid circle indicates cyclotron motion in flat spacetime $m=0$, centred at $r_0=9$, and $\phi_0=0$. The dotted curve shows the drifting of the cyclotron circles for the case $m=1$, $\ell=1$.}
 \label{fig_CyclotronDrift}
\end{figure}

\textbf{Cyclotron drift.} When $m=0$, the situation reduces to a uniform magnetic field in flat spacetime. In this case we find that $\Omega_\phi=0$ and if we further suppose $b=0$, then the solution \Eqref{perturb_soln} describes a circular motion of radius $a$ centred at $r=r_0$, $\phi=\phi_0$ with cyclotron/Larmor frequency \cite{Frolov:2014zia,Kolos:2015iva} 
\begin{align}
 \omega_{r\pm}=\omega_L=2\beta.
\end{align} 
When $m\neq0$, the first term $\Omega_\phi\tau$ in Eq.~\Eqref{perturb_phi_soln} contributes to a linearly increasing azimuthal angle to the harmonic oscillation of the second term. This gives an interpretation of a \emph{drifting} of cyclotron motion due to gravity. An example of this is depicted in Fig.~\ref{fig_CyclotronDrift}, for the case $\beta=0.5$, $r_0=9$, $a=0.1$, $\ell=1$.

\section{Non-equatorial motion} \label{sec_NonEquatorial}

As mentioned in Sec.~\ref{sec_Maxwell}, the behaviour of the test magnetic field fall under two regimes, namely $r<r_c$ and $r>r_c$. For the Hayward black hole case, the magnetic field exterior to the horizon consists of mostly uniform field lines along the $z$-axis, and therefore its off-equator motion of charged particles typically resemble that of the weakly magnetised Schwarzschild case. On the other hand, for the horizonless Hayward spacetime, it is possible for charged particles to access the region $r<r_c$ where the magnetic field resembles the loops of an isolated dipole. As such most of our examples will be in this more interesting situation.

\subsection{Bounce and drift motion}

Because the magnetic field lines take the dipole loop form for $r<r_c$, we might expect the trajectories of charged particles near the core of the horizonless Hayward spacetime to be somewhat analogous to charged particles in planetary magnetic fields. The motion of classical charged particles around planetary magnetic field dipoles have been studied in Refs.~\cite{NorthropBook,WaltBook,McGuire2003,Ozturk:2011}. In particular, the motion of a charged particles around a dipole consists of a so-called \emph{bounce motion} and \emph{drift motion}. 

As the dipole structure is responsible for this bounce and drift behaviour, this should occur for particles in the regime $r\lesssim\brac{4m\ell^2}^{1/3}$. For non-equatorial motion of a particle with a charge $\beta$, energy $E$, and angular momentum $L$, we can check its domain of existence by ensuring the right-hand side of Eq.~\Eqref{eff_energy_eqn},  namely $\mathcal{E}-\mathcal{U}$, is positive. We copy over the expression for $\mathcal{U}$ to this page for the convenience of the reader:
\begin{align*}
 \mathcal{U}=f\sbrac{r^2\sin^2\theta\brac{\frac{L}{r^2\sin^2\theta}-e\chi}^2+1}.
\end{align*}
Recall that $\chi$ becomes negative in the dipole regime $r<\brac{4m\ell^2}^{1/3}$. Therefore, due to the term $\brac{\frac{L}{r^2\sin^2\theta}-e\chi}^2$, the effective potential becomes repulsive if $L$ is positive. This is alleviated for negative $L$, though eventually a potential barrier is still inevitable due to the $1/r^2$ factor. In any case, we see that orbits with $L<0$ can exist closer to the origin where the magnetic field more resembles the dipole. Furthermore, we also note that $\mathcal{U}$ becomes large as $\theta$ approaches the north and south poles. This gives potential barriers for $\theta$ to oscillate between, giving the `bounce' motion.


The `drift' motion can be understood as the extension of the \emph{cyclotron drift} discussed in Sec.~\ref{subsec_circular_perturb} off the equatorial plane. The physics of the drift is similar: The charged particles tend to execute circular motion around magnetic field lines. Here the polar component of the motion carries the cyclotron circle off the equator. At the same time, the radial gravitational and Lorentz forces causes the drift of the cyclotron circle in the $\phi$-direction.

\begin{figure}
 \begin{subfigure}[b]{\textwidth}
    \centering
    \includegraphics[scale=0.4]{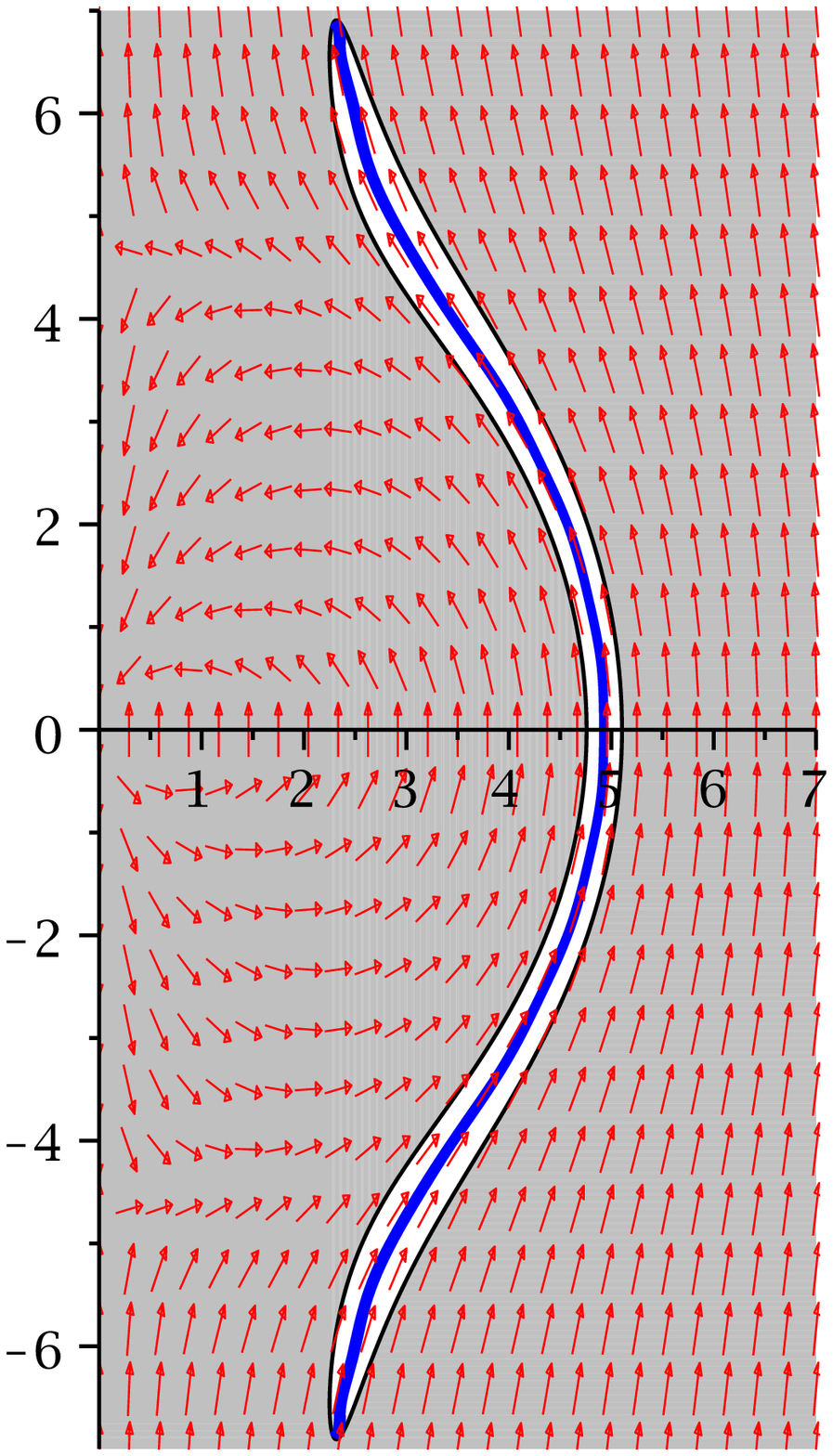}
    \includegraphics[width=0.5\textwidth]{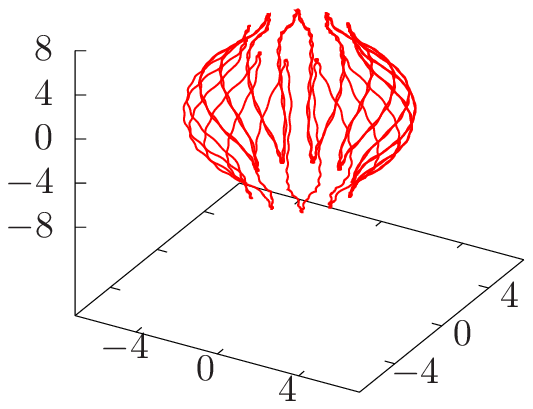}
    \caption{$L=2$, and $r_{\mathrm{init}}=4.9$.}
    \label{fig_Ex1a}
  \end{subfigure}
  \begin{subfigure}[b]{\textwidth}
    \centering
    \includegraphics[scale=0.4]{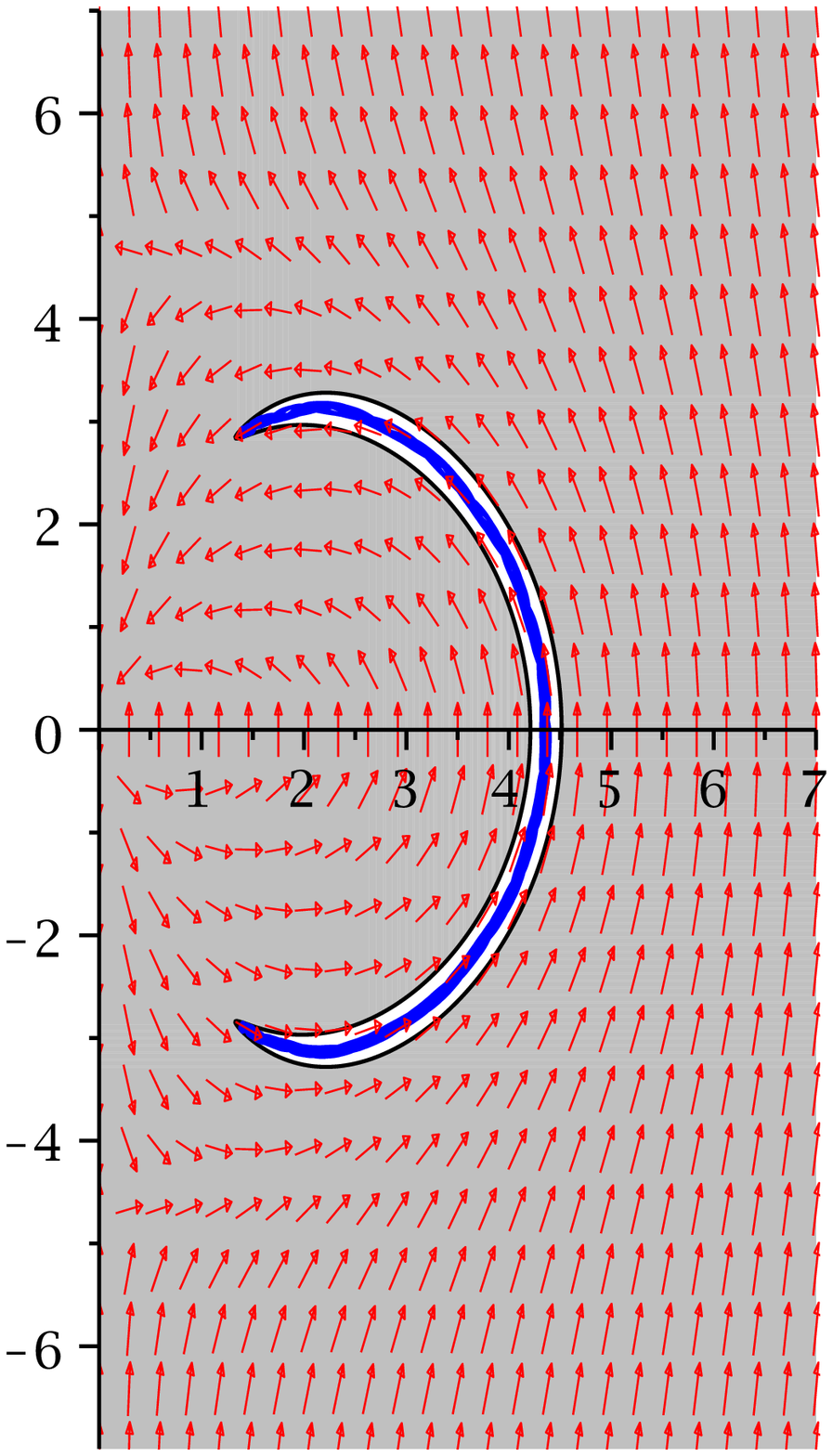}
    \includegraphics[width=0.5\textwidth]{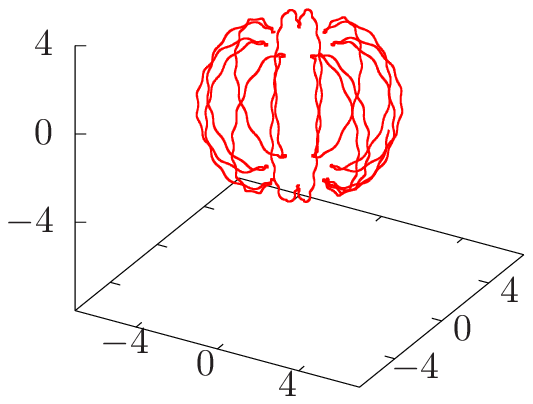}
    \caption{$L=-2$, and $r_{\mathrm{init}}=4.4$.}
    \label{fig_Ex1b}
  \end{subfigure}
  \caption{Orbits for the magnetised horizonless Hayward spacetime with $m=1$, $\ell=5$. The particles have energy $E=0.87$, charge parameter $\beta=0.5$, and are initiated at $\theta=\frac{\pi}{2}$, $r=r_{\mathrm{init}}$, $\dot{r}=0$, with the initial $\dot{\theta}$ determined from Eq.~\Eqref{FirstIntegral}. The left panels are the two-dimensional projections of the space (see main text), with the shaded regions are domains where $\mathcal{E}-\mathcal{U}<0$, which is inaccessible to the particle. The little red arrows indicate the direction of the magnetic field $\vec{B}$ at each point. The blue curve is the trajectory of the particle in the projected coordinates. The right panel shows the same trajectory in the usual three-dimensional Cartesian coordinates.}
  \label{fig_Ex1}
\end{figure}

Turning to numerical solutions to demonstrate the above qualitative reasoning, let us now consider an example in a spacetime of $m=1$, $\ell=5$ with a particle of energy $E=0.87$ with charge parameter $\beta=0.5$ is shown in Fig.~\ref{fig_Ex1}. Angular momenta of two opposite signs are considered; specifically $L=2$ in Fig.~\ref{fig_Ex1a} and $L=-2$ in Fig.~\ref{fig_Ex1b}. In the left panel of each figure is a depiction of the space where the azimuthal direction is projected out. More precisely, it is the map
\begin{align}
 (r,\theta,\phi)\mapsto (r,\theta),
\end{align}
followed by a transformation to cylindrical coordinates with 
\begin{align}
 \rho=r\sin\theta,\quad z=r\cos\theta.
\end{align}
(Note that we have already used these coordinates when plotting $\vec{B}$ in Sec.~\ref{sec_Maxwell}.) The shaded domain correspond to $\mathcal{E}-\mathcal{U}<0$, which is inaccessible to the particle. The particles can only access the white domains where $\mathcal{E}-\mathcal{U}\geq 0$, where the inequality is saturated at the boundary of the domain. The little red arrows on the left panels show the magnetic vector field $\vec{B}$. The blue curve is the trajectory of the particle in these projected coordinates. As expected, it is entirely confined within the accessible (white) domain. The red curve on the right panels are simply the trajectory in the usual three-dimensional Cartesian-type coordinates:
\begin{align}
 X=r\sin\theta\cos\phi,\quad Y=r\sin\theta\sin\phi,\quad Z=r\cos\theta.
\end{align}

\begin{figure}
 \begin{subfigure}[b]{\textwidth}
    \centering
    \includegraphics[scale=0.4]{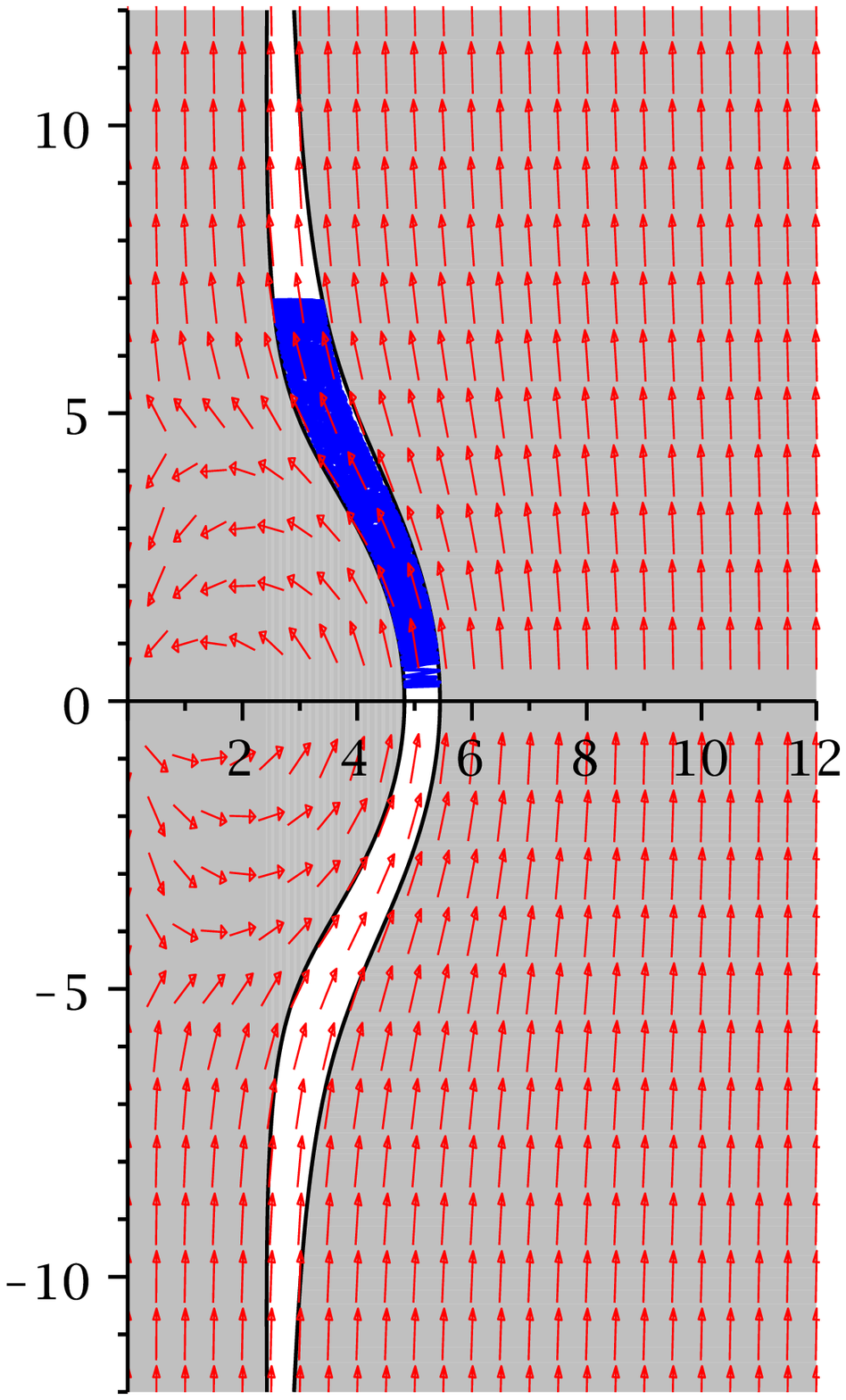}
    \includegraphics[width=0.5\textwidth]{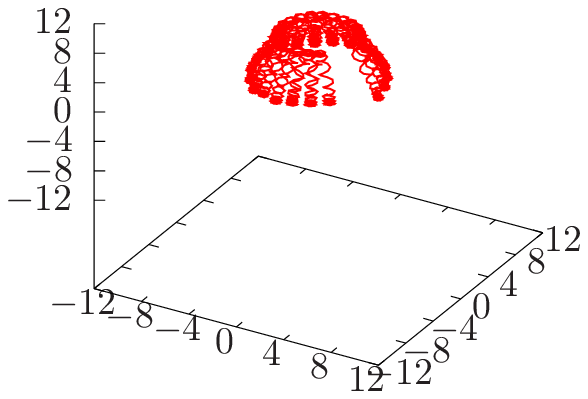}
    \caption{Initial values $\theta_{\mathrm{init}}=\frac{\pi}{2}-0.05$, $\dot{r}=0$, $\dot{\theta}=0$ and $r_{\mathrm{init}}$ determined from Eq.~\Eqref{FirstIntegral}.}
    \label{fig_Ex2a}
  \end{subfigure}
  \begin{subfigure}[b]{\textwidth}
    \centering
    \includegraphics[scale=0.4]{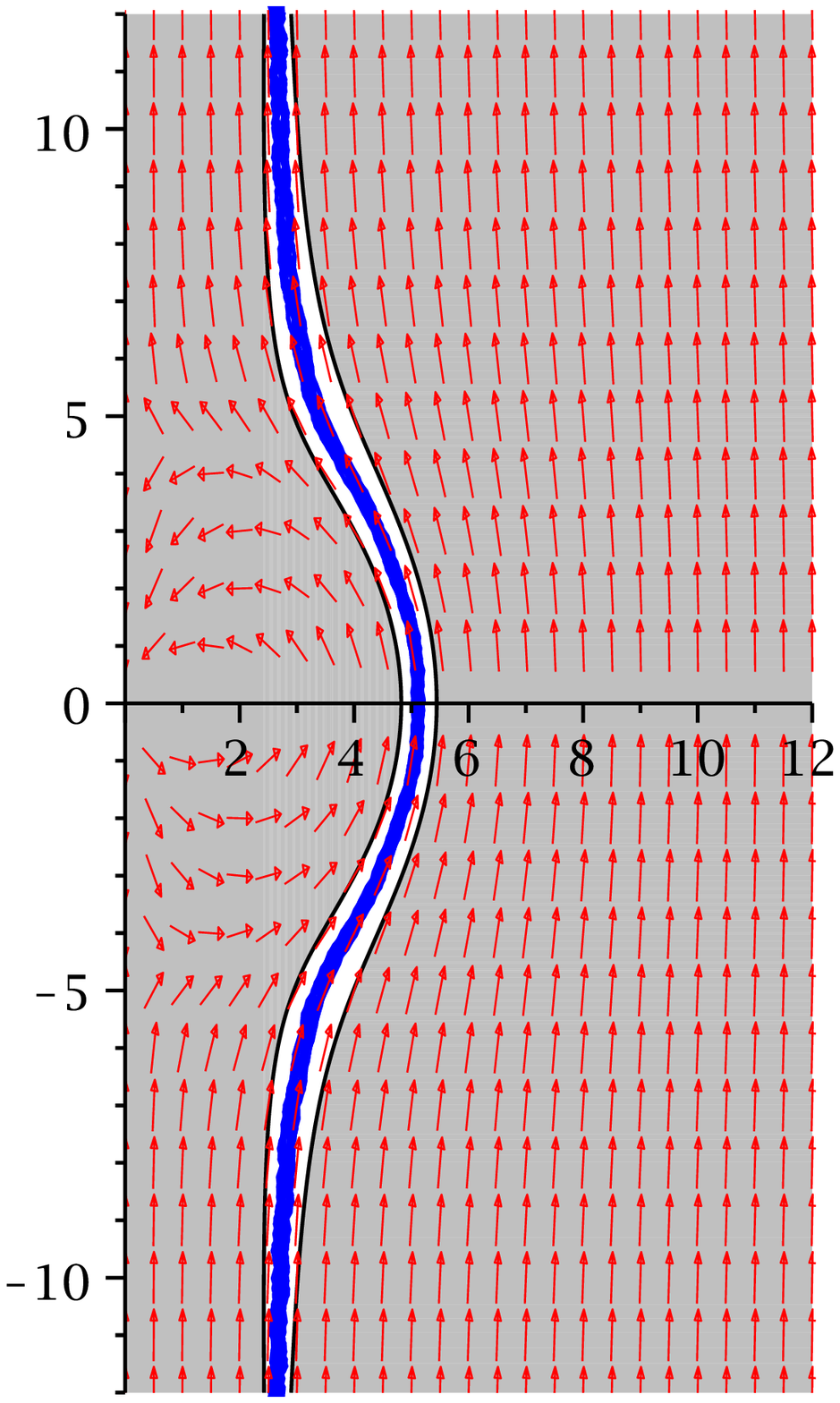}
    \includegraphics[width=0.5\textwidth]{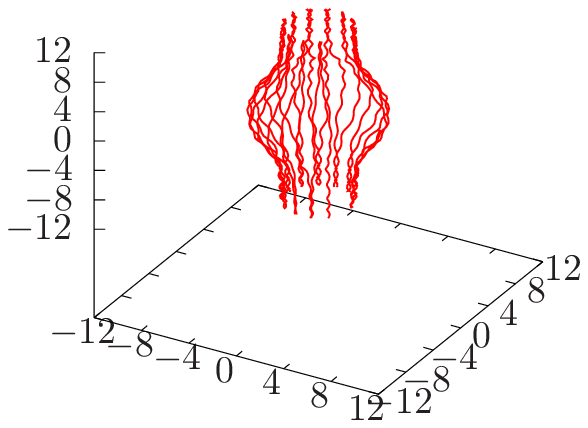}
    \caption{Initial values $\theta_{\mathrm{init}}=\frac{\pi}{2}$, $\dot{r}=0$, and $r_{\mathrm{init}}=5.2$, and $\dot{\theta}$ determined from Eq.~\Eqref{FirstIntegral}.}
    \label{fig_Ex2b}
  \end{subfigure}
  \caption{Orbits for the magnetised horizonless Hayward spacetime with $m=1$, $\ell=5$. The particles with $E=0.95$, $L=4$, and $\beta=0.6$. In Fig.~\ref{fig_Ex2a} the particle starts from rest at a point off the equator, and in Fig.~\ref{fig_Ex2b} the particle starts on the equator with an initial non-zero $\dot{\theta}$.}
  \label{fig_Ex2}
\end{figure}

While the equation $\mathcal{E}-\mathcal{U}\geq0$ depicts the regions accessible to the particle, it may turn out that the particle spends most of its proper time in a subset of this domain, depending on its initial conditions. To illustrate this point, consider Fig.~\ref{fig_Ex2} which shows two orbits of particles with the same $\beta$, $L$, and $E$ in the same spacetime. Only the initial conditions are different. In Fig.~\ref{fig_Ex2a}, the particle starts from rest ($\dot{r}=\dot{\theta}=0$) at a position slightly north of the equator. (Specifically $\theta=\frac{\pi}{2}-0.05$.) We see that the trajectory mostly in the northern hemisphere. On the other hand, in Fig.~\ref{fig_Ex2b} the particle starts at the equator with a non-zero polar velocity ($\dot{\theta}\neq0$, whose specific value must obey Eq.~\Eqref{FirstIntegral} for a given $r=4.9$ and $\theta=\frac{\pi}{2}$). We see that the resulting motion is mostly symmetric about the equator.

\subsection{Polar orbits}

We now turn to the case $L=0$. Here, there are no more terms of the form $1/\sin\theta$ and therefore the particles are able to reach the polar axis at $\theta=0,\pi$. On the axis itself, the $\theta$ equation \Eqref{ddot_theta} on the axis becomes 
\begin{align}
 \frac{1}{r^2}\frac{\dif}{\dif\tau}\brac{r^2\dot{\theta}}=-e^2\chi^2\sin\theta\cos\theta.
\end{align}
We observe that a constant $\theta=0$ or $\pi$ remains a solution. This describes particles of zero angular momentum moving radially along the north or south polar axis. 

Along the axis, the effective potential simplifies to
\begin{align}
 \mathcal{U}(r,0)=\mathcal{U}(0,\pi)&=f(r), 
\end{align}
in particular it becomes independent of $\beta$. This is expected because on the axis itself, $\vec{B}$ points parallel to the axis. So particles moving radially will not experience a Lorentz interaction. Its motion is purely due to the gravitational influence of the Hayward spacetime and would occur for geodesics of neutral particles and/or charged particles with the magnetic field turned off. Indeed, in the horizonless case, the potential has a stable minimum at $r=r_c=\brac{4m\ell^2}^{1/3}$, where
\begin{align}
 \mathcal{U}'(r_c,0)=\mathcal{U}'(r_c,\pi)=\frac{2}{3\ell^2}.
\end{align}

Along the polar axis, we have
\begin{align}
 \lim_{r\rightarrow\infty}\mathcal{U}(r,0)=\lim_{r\rightarrow\infty}\mathcal{U}(r,\pi)=1, 
\end{align}
so particles can escape to infinity along the axis. At the origin, we also have $\mathcal{U}(0,0)=\mathcal{U}(0,\pi)=1$. Therefore unbound particles from infinity along, say, the north pole passes over the potential well and continues passing through the origin, continuing over the southern well and continues to infinity.
 
\begin{figure}
    \centering
    \includegraphics[scale=0.4]{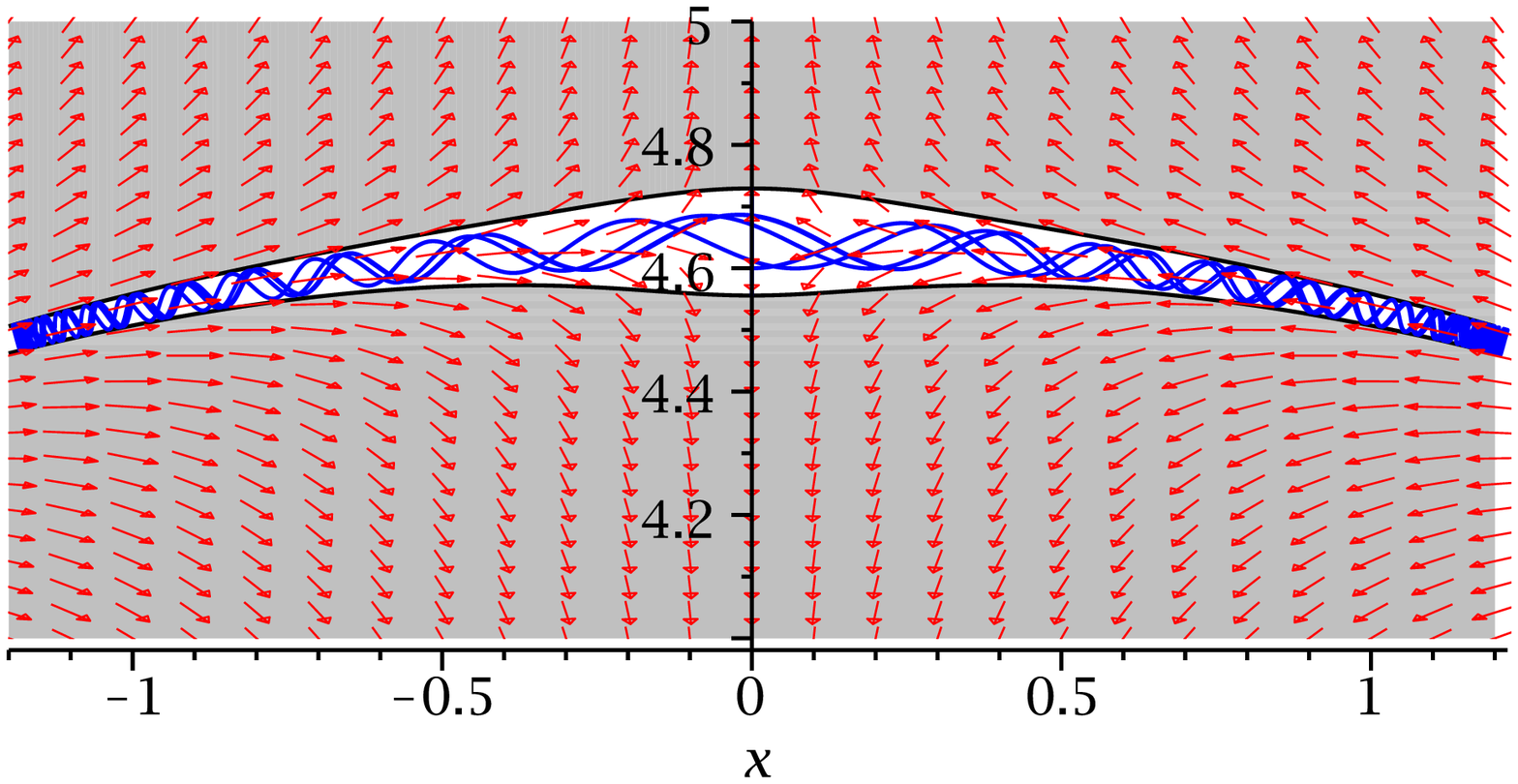}
    \includegraphics[width=0.5\textwidth]{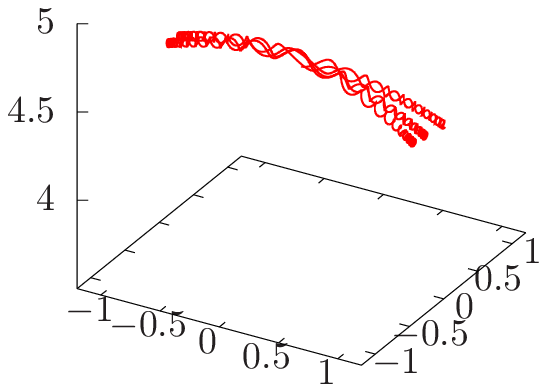}
  \caption{Polar orbits for the magnetised horizonless Hayward spacetime with $m=1$, $\ell=5$. The particles have energy $E=0.8443$, and zero angular momentum, $L=0$. The particle has charge parameter $\beta=0.7$, and is initiated at $\theta=0$, $r=4.6$, $\dot{r}=0$, with the initial $\dot{\theta}$ determined from Eq.~\Eqref{FirstIntegral}.}
  \label{fig_Ex3a}
\end{figure}

Particles with energies $\mathcal{E}=\mathcal{U}(r_c,0)$ or $\mathcal{E}=\mathcal{U}(r_c,\pi)$ experience two stable minimum potential at $r=r_c$, one for each $\theta=0$ and $\theta=\pi$, respectively. Increasing the energy slightly higher than $\mathcal{U}(r_c,0)$ or $\mathcal{U}(r_c,\pi)$ leads to a finite potential well that bounds the particle there. If a particle starts along the axis with $\dot{\theta}=0$, it oscillates about $r_c$ in a straight line along the axis. Otherwise, giving it a slight polar velocity ($\dot{\theta}\neq0$), it will then follow a more complicated curved trajectory around the potential well. For certain ranges of energy, the particle is trapped in a well that is in one hemisphere containing $r_c$, never crossing the equator. An example of this is shown in Fig.~\ref{fig_Ex3a}, where the motion is plotted for a particle of $\beta=0.7$ and energy $E=0.8443$ in a spacetime of $m=1$ and $\ell=5$.

\begin{figure}
    \centering
    \includegraphics[scale=0.4]{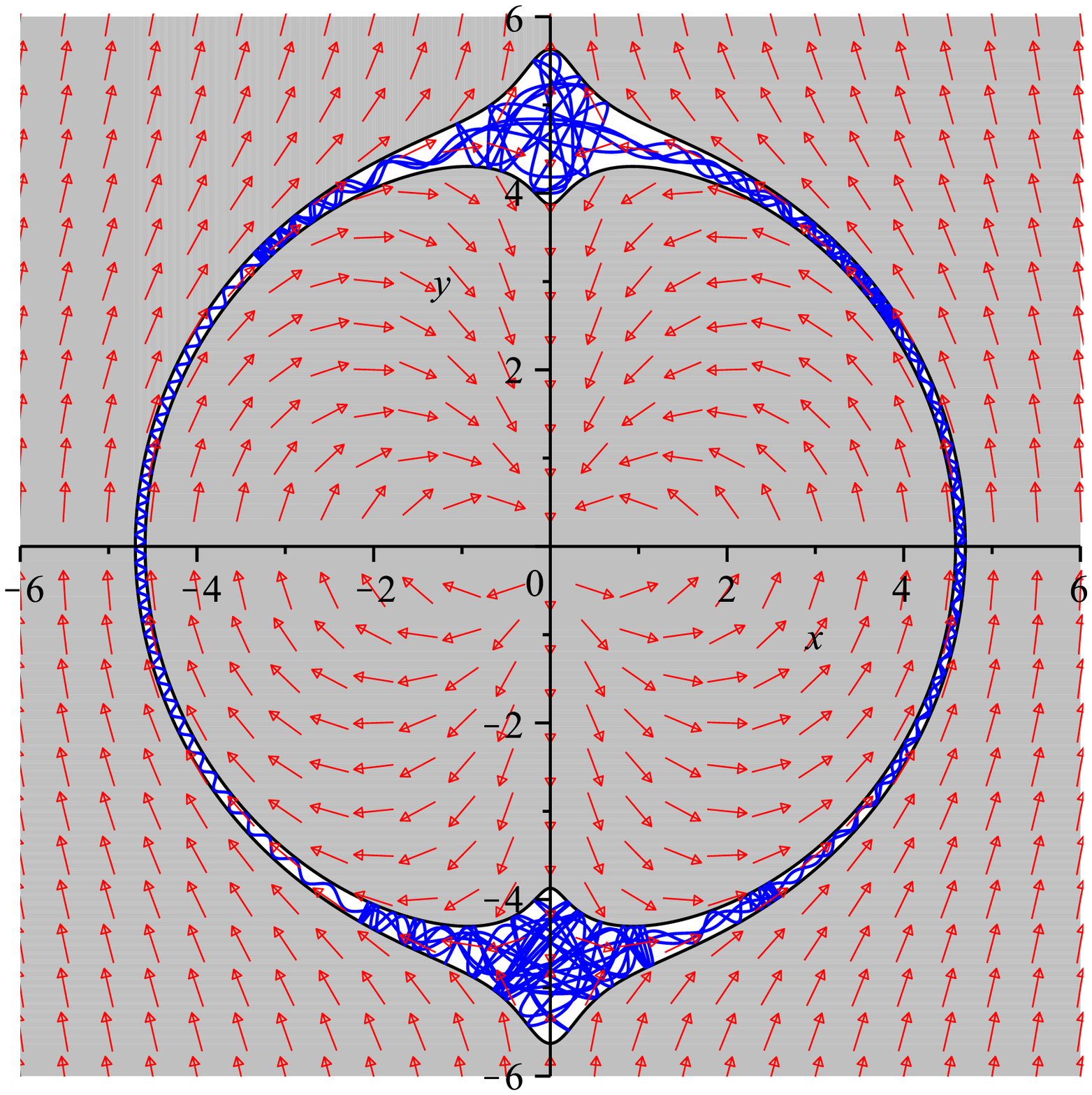}
    \includegraphics[width=0.5\textwidth]{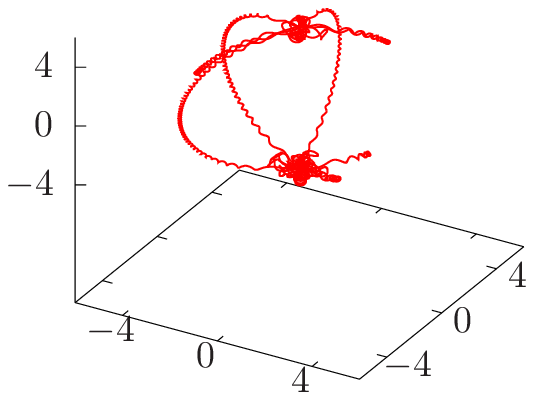}
  \caption{Polar orbits for the magnetised horizonless Hayward spacetime with $m=1$, $\ell=5$. The particles have energy $E=0.85$, and zero angular momentum, $L=0$. The particle has charge parameter $\beta=0.5$, and is initiated at $\theta=0$, $r=4$, $\dot{r}=0$, with the initial $\dot{\theta}$ determined from Eq.~\Eqref{FirstIntegral}.}
  \label{fig_Ex3b}
\end{figure}

If we increase the energy further, the potential wells in the north and south grows and connects with each other via a narrow path across the equator. With the right initial conditions, a charged particle may, for instance, start near $r_c$ at the north, and gets bounced around in the northern potential well. Eventually it may happen to be bounced into the narrow path southwards toward the equator and spends some time in the southern well. This can proceed back and forth. An example of this is shown in Fig.~\ref{fig_Ex3b}, for $\beta=0.5$, $E=0.85$ in a spacetime $m=1$ and $\ell=5$.

\section{Conclusion} \label{sec_conclusion}

In this work, we have immersed a Hayward spacetime in an external magnetic field. This was achieved by solving Maxwell's equations with the Hayward geometry as a fixed background. A crucial requirement of this solution is that the magnetic field strengths are \emph{test} fields --- they are sufficiently weak so that fields do not backreact to the spacetime curvature, and that the geometry continues to be described by the Hayward metric.

The structure of the magnetic field depends on the characteristic radius $r_c=\brac{4m\ell^2}^{1/3}$. At $r\gtrsim r_c$, the magnetic field tends to be homogeneous and parallel to the $z$-axis, in agreement to the weakly magnetised Schwarzschild solution. For $r\lesssim r_c$, the magnetic field lines have a dipole-loop structure. There are two points where the field is zero, which is at $r=r_c$ on the north and south $z$-axis. 

The equations of motion for a charged particle was derived using the Lagrangian formalism. For magnetised Hayward black holes, particles behave mostly similar to the magnetised Schwarzschild case, with the Hayward parameter $\ell$ modifying the details of the orbital parameters. In particular, there are two sets of circular orbits with energy and angular momentum are given by $E_\pm$ and $L_\pm$, respectively. The presence of $\ell$ decreases the size of the ISCOs (innermost stable circular orbits). For the horizonless Hayward case, there exists a second potential well near the origin. Therefore, within each set of circular orbit (the `$+$' or `$-$' set), there are generically three branches of circular orbits, two of which are stable. We have identified two radii $r_{\mathrm{OCCO}\pm}$ and $r_{\mathrm{ICCO}\pm}$, where circular orbits of radii $r_{\mathrm{ICCO}\pm}<r<r_{\mathrm{OCCO}\pm}$ are unstable. The two stable branches are circular orbits of radii $r<r_{\mathrm{ICCO}\pm}$ and $r>r_{\mathrm{OCCO}\pm}$.

Due to the looped field lines of $r<r_c$, charged particles in this region behave differently from that of the magnetised Schwarzschild case. For certain choices of parameters, the charged particles move in helical motion around a field line, bouncing between two fixed polar angles. This bounce motion additionally has a drift in the azimuthal direction. So the motion can be seen as analogous to that of charged particles around planetary magnetic fields.

\section*{Acknowledgments}
 
Y.-K.~L is supported by Xiamen University Malaysia Research Fund (Grant no. XMUMRF/ 2021-C8/IPHY/0001).

\bibliographystyle{mHay}

\bibliography{mHay}

\providecommand{\href}[2]{#2}\begingroup\raggedright\begin{thebibliography}{10}

\bibitem{Bardeen}
J.~M. Bardeen Proceedings of GR5, (Tiflis, Georgia), 1968.

\bibitem{Dymnikova:1992ux}
I.~Dymnikova, {\it {Vacuum nonsingular black hole}},  Gen. Rel. Grav. {\bf 24}
  (1992) 235--242.

\bibitem{Ayon-Beato:1999qin}
E.~Ayon-Beato and A.~Garcia, {\it {Nonsingular charged black hole solution for
  nonlinear source}},  Gen. Rel. Grav. {\bf 31} (1999) 629--633,
  [\href{http://arxiv.org/abs/gr-qc/9911084}{{\tt gr-qc/9911084}}].

\bibitem{Ansoldi:2008jw}
S.~Ansoldi, {\it {Spherical black holes with regular center: A Review of
  existing models including a recent realization with Gaussian sources}},  in
  {\em {Conference on Black Holes and Naked Singularities}}, 2, 2008.
\newblock \href{http://arxiv.org/abs/0802.0330}{{\tt arXiv:0802.0330}}.

\bibitem{Hayward:2005gi}
S.~A. Hayward, {\it {Formation and evaporation of nonsingular black holes}},
  Phys. Rev. Lett. {\bf 96} (2006) 031103,
  [\href{http://arxiv.org/abs/gr-qc/0506126}{{\tt gr-qc/0506126}}].

\bibitem{Kruglov2015}
S.~I. Kruglov, {\it A model of nonlinear electrodynamics},  Annals of Physics
  {\bf 353} (2015) 299–306.

\bibitem{Ali:2019bcn}
A.~Ali and K.~Saifullah, {\it {Asymptotic magnetically charged non-singular
  black hole and its thermodynamics}},  Phys. Lett. B {\bf 792} (2019)
  276--283, [\href{http://arxiv.org/abs/1904.05727}{{\tt arXiv:1904.05727}}].

\bibitem{Mazharimousavi:2019jja}
S.~H. Mazharimousavi and M.~Halilsoy, {\it {Note on regular magnetic black
  hole}},  Phys. Lett. B {\bf 796} (2019) 123--125.

\bibitem{Kruglov2021}
S.~I. Kruglov, {\it Regular model of magnetized black hole with rational
  nonlinear electrodynamics},  International Journal of Modern Physics A {\bf
  36} (2021), no.~21 2150158.

\bibitem{IlichKruglov:2021pdw}
S.~I. Kruglov, {\it {Remarks on Nonsingular Models of Hayward and Magnetized
  Black Hole with Rational Nonlinear Electrodynamics}},  Grav. Cosmol. {\bf 27}
  (2021), no.~1 78--84, [\href{http://arxiv.org/abs/2103.14087}{{\tt
  arXiv:2103.14087}}].

\bibitem{Flachi:2012nv}
A.~Flachi and J.~P.~S. Lemos, {\it {Quasinormal modes of regular black holes}},
   Phys. Rev. D {\bf 87} (2013), no.~2 024034,
  [\href{http://arxiv.org/abs/1211.6212}{{\tt arXiv:1211.6212}}].

\bibitem{Halilsoy:2013iza}
M.~Halilsoy, A.~Ovgun, and S.~H. Mazharimousavi, {\it {Thin-shell wormholes
  from the regular Hayward black hole}},  Eur. Phys. J. C {\bf 74} (2014) 2796,
  [\href{http://arxiv.org/abs/1312.6665}{{\tt arXiv:1312.6665}}].

\bibitem{Pourhassan:2016qoz}
B.~Pourhassan, M.~Faizal, and U.~Debnath, {\it {Effects of Thermal Fluctuations
  on the Thermodynamics of Modified Hayward Black Hole}},  Eur. Phys. J. C {\bf
  76} (2016), no.~3 145, [\href{http://arxiv.org/abs/1603.01457}{{\tt
  arXiv:1603.01457}}].

\bibitem{Um:2019qgc}
H.~Um and W.~Kim, {\it {Formation of the Hayward black hole from a collapsing
  shell}},  Phys. Rev. D {\bf 101} (2020), no.~6 065017,
  [\href{http://arxiv.org/abs/1912.04490}{{\tt arXiv:1912.04490}}].

\bibitem{NaveenaKumara:2020lgq}
N.~K. et~al., {\it {Microstructure and continuous phase transition of a regular
  Hayward black hole in anti-de Sitter spacetime}},  PTEP {\bf 2021} (2021),
  no.~7 073E01, [\href{http://arxiv.org/abs/2003.00889}{{\tt
  arXiv:2003.00889}}].

\bibitem{Abbas:2014oua}
G.~Abbas and U.~Sabiullah, {\it {Geodesic Study of Regular Hayward Black
  Hole}},  Astrophys. Space Sci. {\bf 352} (2014) 769--774,
  [\href{http://arxiv.org/abs/1406.0840}{{\tt arXiv:1406.0840}}].

\bibitem{Schee:2015nua}
J.~Schee and Z.~Stuchlik, {\it {Gravitational lensing and ghost images in the
  regular Bardeen no-horizon spacetimes}},  JCAP {\bf 06} (2015) 048,
  [\href{http://arxiv.org/abs/1501.00835}{{\tt arXiv:1501.00835}}].

\bibitem{Chiba:2017nml}
T.~Chiba and M.~Kimura, {\it {A note on geodesics in the Hayward metric}},
  PTEP {\bf 2017} (2017), no.~4 043E01,
  [\href{http://arxiv.org/abs/1701.04910}{{\tt arXiv:1701.04910}}].

\bibitem{Hu:2018old}
J.-P. Hu, Y.~Zhang, L.-L. Shi, and P.-F. Duan, {\it {Structure of geodesics in
  the regular Hayward black hole space-time}},  Gen. Rel. Grav. {\bf 50}
  (2018), no.~7 89, [\href{http://arxiv.org/abs/2104.07506}{{\tt
  arXiv:2104.07506}}].

\bibitem{Bambi:2013ufa}
C.~Bambi and L.~Modesto, {\it {Rotating regular black holes}},  Phys. Lett. B
  {\bf 721} (2013) 329--334, [\href{http://arxiv.org/abs/1302.6075}{{\tt
  arXiv:1302.6075}}].

\bibitem{Torres:2016pgk}
R.~Torres and F.~Fayos, {\it {On regular rotating black holes}},  Gen. Rel.
  Grav. {\bf 49} (2017), no.~1 2, [\href{http://arxiv.org/abs/1611.03654}{{\tt
  arXiv:1611.03654}}].

\bibitem{Bautista-Olvera:2019blb}
B.~Bautista-Olvera, J.~C. Degollado, and G.~German, {\it {Geodesic structure of
  a rotating regular black hole}},  \href{http://arxiv.org/abs/1908.01886}{{\tt
  arXiv:1908.01886}}.

\bibitem{Amir:2015pja}
M.~Amir and S.~G. Ghosh, {\it {Rotating Hayward regular black hole as particle
  accelerator}},  JHEP {\bf 07} (2015) 015,
  [\href{http://arxiv.org/abs/1503.08553}{{\tt arXiv:1503.08553}}].

\bibitem{Kumar:2019pjp}
R.~Kumar, S.~G. Ghosh, and A.~Wang, {\it {Shadow cast and deflection of light
  by charged rotating regular black holes}},  Phys. Rev. D {\bf 100} (2019),
  no.~12 124024, [\href{http://arxiv.org/abs/1912.05154}{{\tt
  arXiv:1912.05154}}].

\bibitem{Aliev:1986wu}
A.~N. Aliev, D.~V. Galtsov, and V.~I. Petrukhov, {\it {Negative absorption near
  a magnetized black hole: Black hole masers}},  Astrophys. Space Sci. {\bf
  124} (1986) 137.

\bibitem{Aliev:1989wx}
A.~N. Aliev and D.~V. Galtsov, {\it {Magnetized Black Holes}},  Sov. Phys. Usp.
  {\bf 32} (1989) 75.

\bibitem{Aliev:2002nw}
A.~N. Aliev and N.~Ozdemir, {\it {Motion of charged particles around a rotating
  black hole in a magnetic field}},  Mon. Not. Roy. Astron. Soc. {\bf 336}
  (2002) 241--248, [\href{http://arxiv.org/abs/gr-qc/0208025}{{\tt
  gr-qc/0208025}}].

\bibitem{Frolov:2010mi}
V.~P. Frolov and A.~A. Shoom, {\it {Motion of charged particles near weakly
  magnetized Schwarzschild black hole}},  Phys. Rev. D {\bf 82} (2010) 084034,
  [\href{http://arxiv.org/abs/1008.2985}{{\tt arXiv:1008.2985}}].

\bibitem{Frolov:2011ea}
V.~P. Frolov, {\it {Weakly magnetized black holes as particle accelerators}},
  Phys. Rev. D {\bf 85} (2012) 024020,
  [\href{http://arxiv.org/abs/1110.6274}{{\tt arXiv:1110.6274}}].

\bibitem{Zahrani:2013up}
A.~M.~A. Zahrani, V.~P. Frolov, and A.~A. Shoom, {\it {Critical escape velocity
  for a charged particle moving around a weakly magnetized Schwarzschild black
  hole}},  Phys. Rev. D {\bf 87} (2013), no.~8 084043,
  [\href{http://arxiv.org/abs/1301.4633}{{\tt arXiv:1301.4633}}].

\bibitem{Shiose:2014bqa}
R.~Shiose, M.~Kimura, and T.~Chiba, {\it {Motion of Charged Particles around a
  Weakly Magnetized Rotating Black Hole}},  Phys. Rev. D {\bf 90} (2014),
  no.~12 124016, [\href{http://arxiv.org/abs/1409.3310}{{\tt
  arXiv:1409.3310}}].

\bibitem{Kolos:2015iva}
M.~Kolo\v{s}, Z.~Stuchl\'\i{}k, and A.~Tursunov, {\it {Quasi-harmonic
  oscillatory motion of charged particles around a Schwarzschild black hole
  immersed in a uniform magnetic field}},  Class. Quant. Grav. {\bf 32} (2015),
  no.~16 165009, [\href{http://arxiv.org/abs/1506.06799}{{\tt
  arXiv:1506.06799}}].

\bibitem{Shoom:2015uba}
A.~A. Shoom, {\it {Synchrotron radiation from a weakly magnetized Schwarzschild
  black hole}},  Phys. Rev. D {\bf 92} (2015), no.~12 124066,
  [\href{http://arxiv.org/abs/1509.02535}{{\tt arXiv:1509.02535}}].

\bibitem{Herdeiro:2015vaa}
C.~Herdeiro and E.~Radu, {\it {Anti-de-Sitter regular electric multipoles:
  Towards Einstein\textendash{}Maxwell-AdS solitons}},  Phys. Lett. B {\bf 749}
  (2015) 393--398, [\href{http://arxiv.org/abs/1507.04370}{{\tt
  arXiv:1507.04370}}].

\bibitem{Costa:2015gol}
M.~S. Costa, L.~Greenspan, M.~Oliveira, J.~a. Penedones, and J.~E. Santos, {\it
  {Polarised Black Holes in AdS}},  Class. Quant. Grav. {\bf 33} (2016), no.~11
  115011, [\href{http://arxiv.org/abs/1511.08505}{{\tt arXiv:1511.08505}}].

\bibitem{Vesely:2019ajp}
J.~Vesel\'y and M.~\v{Z}ofka, {\it {Cosmological magnetic field: The
  boost-symmetric case}},  Phys. Rev. D {\bf 100} (2019), no.~4 044059,
  [\href{http://arxiv.org/abs/2104.02123}{{\tt arXiv:2104.02123}}].

\bibitem{Vesely:2021jlc}
J.~Vesel\'y and M.~\v{Z}ofka, {\it {Cylindrical spacetimes due to radial
  magnetic fields}},  Phys. Rev. D {\bf 103} (2021), no.~2 024048,
  [\href{http://arxiv.org/abs/2104.01557}{{\tt arXiv:2104.01557}}].

\bibitem{Wald:1974np}
R.~M. Wald, {\it {Black hole in a uniform magnetic field}},  Phys. Rev. D {\bf
  10} (1974) 1680--1685.

\bibitem{Azreg-Ainou:2016tkt}
M.~Azreg-A\"\i{}nou, {\it {Vacuum and nonvacuum black holes in a uniform
  magnetic field}},  Eur. Phys. J. C {\bf 76} (2016), no.~7 414,
  [\href{http://arxiv.org/abs/1603.07894}{{\tt arXiv:1603.07894}}].

\bibitem{Petterson:1974bt}
J.~A. Petterson, {\it {Magnetic field of a current loop around a Schwarzschild
  black hole}},  Phys. Rev. D {\bf 10} (1974) 3166--3170.

\bibitem{Lim:2020fnx}
Y.-K. Lim, {\it {Hypocycloid motion in the Melvin magnetic universe}},  Phys.
  Rev. D {\bf 101} (2020), no.~10 104031,
  [\href{http://arxiv.org/abs/2004.08027}{{\tt arXiv:2004.08027}}].

\bibitem{Frolov:2014zia}
V.~P. Frolov, A.~A. Shoom, and C.~Tzounis, {\it {Spectral line broadening in
  magnetized black holes}},  JCAP {\bf 07} (2014) 059,
  [\href{http://arxiv.org/abs/1405.0510}{{\tt arXiv:1405.0510}}].

\bibitem{NorthropBook}
{T.~Northrop}, {\em {The Adiabatic Motion of Charged Particles}}.
\newblock {John Wiley \& Sons}, {New York}, (1963).

\bibitem{WaltBook}
{M.~Walt}, {\em {Introduction to Geomagnetically Trapped Radiation}}.
\newblock {Cambridge University Press}, {Cambridge}, (1994).

\bibitem{McGuire2003}
{G.~C.~McGuire}, {\it {Using computer algebra to investigate the motion of an
  electric charge in magnetic and electric dipole fields}},  Am.~J.~Phys. {\bf
  71} (2003) 809.

\bibitem{Ozturk:2011}
{\"{O}zt\"{u}rk, M.~Kaan}, {\it {Trajectories of charged particles trapped in
  Earth's magnetic field}},  Am.~J.~Phys. {\bf 80} (2012) 420,
  [\href{http://arxiv.org/abs/1112.3487}{{\tt arXiv:1112.3487}}].

\end{thebibliography}\endgroup

\end{document}